\documentclass{emulateapj}

\usepackage{amsmath, amssymb, comment, graphicx, multirow, xspace}
\usepackage{makecell}
\usepackage{longtable}
\usepackage{array} 
\usepackage{academicons} 
\usepackage{xcolor} 
\usepackage{scalerel}
\usepackage{tikz}
\usetikzlibrary{svg.path}
\definecolor{orcidlogocol}{HTML}{A6CE39}
\tikzset{
  orcidlogo/.pic={
    \fill[orcidlogocol] svg{M256,128c0,70.7-57.3,128-128,128C57.3,256,0,198.7,0,128C0,57.3,57.3,0,128,0C198.7,0,256,57.3,256,128z};
    \fill[white] svg{M86.3,186.2H70.9V79.1h15.4v48.4V186.2z}
                 svg{M108.9,79.1h41.6c39.6,0,57,28.3,57,53.6c0,27.5-21.5,53.6-56.8,53.6h-41.8V79.1z M124.3,172.4h24.5c34.9,0,42.9-26.5,42.9-39.7c0-21.5-13.7-39.7-43.7-39.7h-23.7V172.4z}
                 svg{M88.7,56.8c0,5.5-4.5,10.1-10.1,10.1c-5.6,0-10.1-4.6-10.1-10.1c0-5.6,4.5-10.1,10.1-10.1C84.2,46.7,88.7,51.3,88.7,56.8z};
  }
}

\usepackage[encapsulated]{CJK} 
\usepackage{ucs}
\usepackage[utf8x]{inputenc}
\newcommand{\cntext}[1]{\begin{CJK}{UTF8}{gbsn}#1\end{CJK}}

\usepackage[english]{babel}
\usepackage[bookmarks=false,         
            pdfnewwindow=true,      
            colorlinks=true,    
            linkcolor=blue,     
            citecolor=blue,     
            filecolor=blue,  
            urlcolor=blue,      
            ]{hyperref}

\addto\captionsenglish{%
}
\addto\extrasenglish{%
}

\usepackage{natbib}
\usepackage[titletoc,title]{appendix}

\usepackage{lineno}

\newcommand{\noprint}[1]{}
\newcommand{\figsetstart}{{\bf Fig. Set} }
\newcommand{\figsetend}{}
\newcommand{\figsetgrpstart}{}
\newcommand{\figsetgrpend}{}
\newcommand{\figsetnum}[1]{{\bf #1.}}
\newcommand{\figsettitle}[1]{ {\bf #1} }
\newcommand{\figsetgrpnum}[1]{\noprint{#1}}
\newcommand{\figsetgrptitle}[1]{\noprint{#1}}
\newcommand{\figsetplot}[1]{\noprint{#1}}
\newcommand{\figsetgrpnote}[1]{\noprint{#1}}

\newcommand{\Fermi}{$Fermi$\xspace}
\newcommand{\Pvalue}{Poissonian probability\xspace}
\newcommand{\pvalue}{$p$-value\xspace}

\newcommand\colorAutoref[1]{{\hypersetup{linkcolor=blue}\autoref{#1}}}  
\newcommand{\lz}[1]{#1} 

\newcommand\orcidicon[1]{\href{https://orcid.org/#1}{\mbox{\scalerel*{
\begin{tikzpicture}[yscale=-1,transform shape]
\pic{orcidlogo};
\end{tikzpicture}
}{|}}}}

\def\PhysUCSB{1}
\def\PhysUIUC{2}
\def\AstrUIUC{3}
\def\CAPS{4}
\def\Clemson{5}
\def\AstrphysUCLA{6}
\def\CWRU{7}
\def\KIPAC{8}

\begin{document}

\title{New identifications and multi-wavelength properties of extragalactic Fermi Gamma-Ray sources in the SPT-SZ survey field}

\author{
	Lizhong~Zhang~\cntext{(张力中)}\orcidicon{0000-0003-0232-0879},\altaffilmark{\PhysUCSB, \PhysUIUC, \AstrUIUC}
	Joaquin~D.~Vieira\orcidicon{0000-0001-7192-3871},\altaffilmark{\PhysUIUC, \AstrUIUC, \CAPS}
	Marco~Ajello\orcidicon{0000-0002-6584-1703},\altaffilmark{\Clemson}
	Matthew~A.~Malkan\orcidicon{0000-0001-6919-1237},\altaffilmark{\AstrphysUCLA}
	Melanie~A.~Archipley\orcidicon{0000-0002-0517-9842},\altaffilmark{\AstrUIUC, \CAPS}
	Joseph~Capota\orcidicon{0000-0002-8491-3012},\altaffilmark{\PhysUIUC}
	Allen~Foster\orcidicon{0000-0002-7145-1824},\altaffilmark{\CWRU}
	Greg~Madejski\altaffilmark{\KIPAC}
}

\altaffiltext{\PhysUCSB}{Department of Physics, University of California, Santa Barbara, CA 93106, USA}
\altaffiltext{\PhysUIUC}{Department of Physics, University of Illinois at Urbana-Champaign, 1110 West Green St, Urbana, IL 61801, USA}
\altaffiltext{\AstrUIUC}{Department of Astronomy, University of Illinois at Urbana-Champaign, 1002 West Green St, Urbana, IL 61801, USA}
\altaffiltext{\CAPS}{Center for AstroPhysical Surveys, National Center for Supercomputing Applications, Urbana, IL, 61801, USA}
\altaffiltext{\Clemson}{Department of Physics and Astronomy, Clemson University, 118 Kinard Laboratory, Clemson, SC 29631, USA}
\altaffiltext{\AstrphysUCLA}{Department of Physics and Astronomy, University of California, Los Angeles, 475 Portola Plaza, Los Angeles, CA 90095-1547, USA}
\altaffiltext{\CWRU}{Department of Physics, Case Western Reserve University, Cleveland, OH, 44106, USA}
\altaffiltext{\KIPAC}{Kavli Institute for Particle Astrophysics and Cosmology, Department of Physics and SLAC National Accelerator Laboratory, Stanford University, Stanford, CA 94305, USA}

\begin{abstract}
The fourth \Fermi Large Area Telescope (LAT) catalog (4FGL) contains 5064 $\gamma$-ray sources detected at high significance, but 26\% of them still lack associations at other wavelengths.  The SPT-SZ survey, conducted between 2008 and 2011 with the South Pole Telescope (SPT), covers 2500 $\mathrm{deg^2}$ of the Southern sky in three millimeter-wavelength (mm) bands and was used to construct a catalog of nearly 5000 emissive sources. In this study, we introduce a new cross-matching scheme to search for multi-wavelength counterparts of extragalactic $\gamma$-ray sources using a mm catalog. We apply a \Pvalue to evaluate the rate of spurious false associations and compare the multi-wavelength associations from the radio, mm, near-infrared, and X-ray with 4FGL $\gamma$-ray sources.  In the SPT-SZ survey field, 85\% of 4FGL sources are associated with mm counterparts.  These mm sources include 94\% of previously associated 4FGL sources and 56\% of previously unassociated 4FGL sources. The latter group contains 40 4FGL sources for which SPT has provided the first identified counterparts. Nearly all of the SPT-associated 4FGL sources can be described as flat-spectrum radio quasars or blazars.  We find that the mm band is the most efficient wavelength for detecting $\gamma$-ray blazars when considering both completeness and purity.  We also demonstrate that the mm band correlates better to the $\gamma$-ray band than the radio or X-ray bands.  With the next generation of CMB experiments, this technique can be extended to greater sensitivities and more sky area to further complete the identifications of the remaining unknown $\gamma$-ray blazars. 
\end{abstract}

\bigskip\bigskip

\section{Introduction}
\label{sec:intro}

\begin{figure*}
\centering
\includegraphics[width=1.00\textwidth]{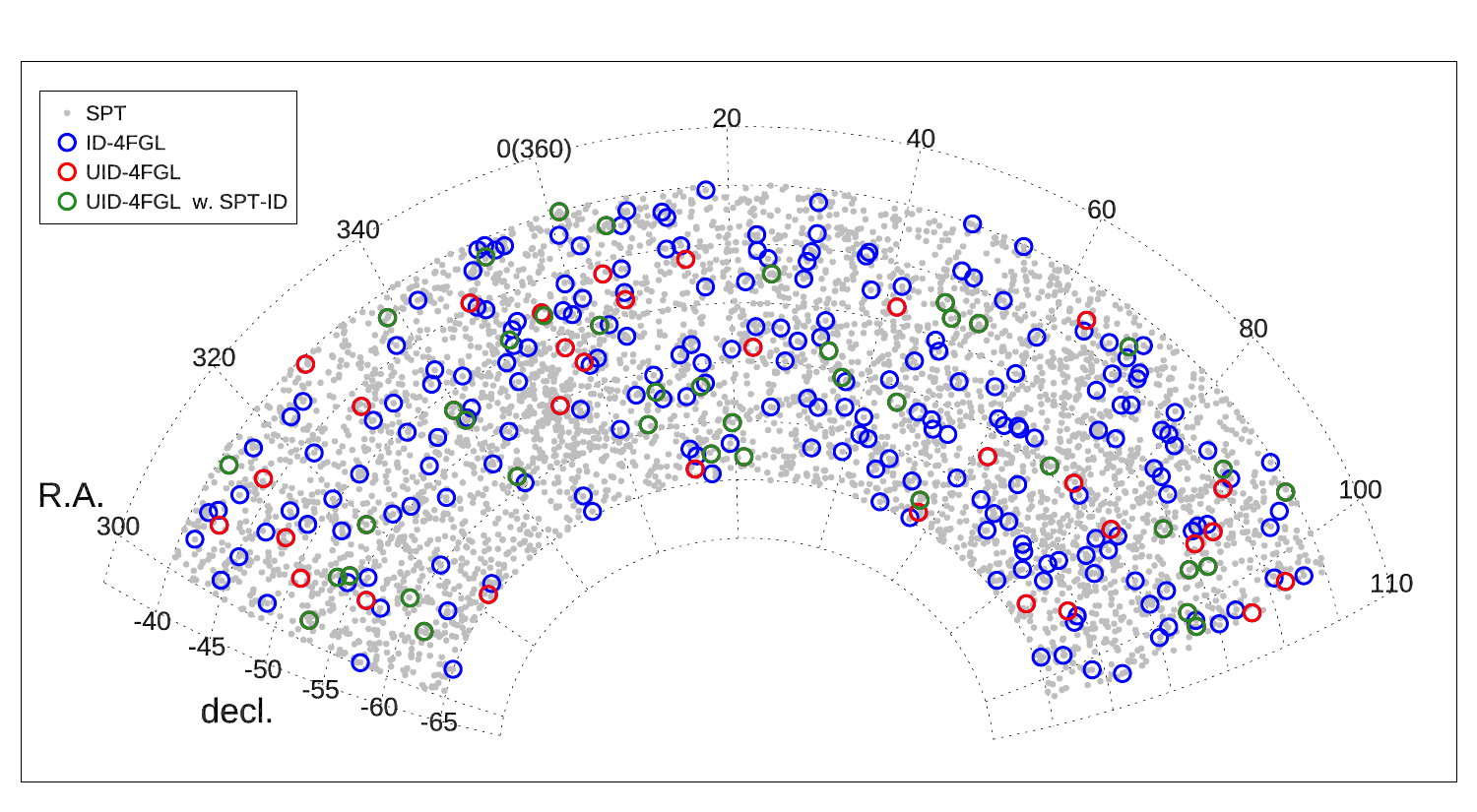}
\caption{The 2500 $\mathrm{deg}^2$ SPT-SZ survey \citep{2020arXiv200303431E}. In this region, \textit{grey points} mark all SPT sources in \citet{2020arXiv200303431E}; \textit{green circles} represent the position of the previously unassociated 4FGL sources which have SPT counterparts (40); \textit{blue circles} represent the already associated 4FGL sources (211); \textit{red circles} represent the remaining still-unidentified 4FGL sources (31).}
\label{fig:fields}
\end{figure*}

The $\gamma$-ray background holds important clues to the nature of galaxy evolution, the cosmic history of black hole accretion, and possibly the nature of dark matter \citep{2015ApJ...800L..27A, 2015ARNPS..65..245F}. Unveiling the nature of the unassociated \Fermi Large Area Telescope (LAT) sources is one of the biggest challenges in $\gamma$-ray astronomy, due to the relatively large point spread function, and is necessary to understand the contribution of various source classes to the $\gamma$-ray background. 

Blazars are a subclass of active galactic nuclei (AGN) with relativistic jets of high-energy particles pointing near our line of sight (e.g., \citealt{1995PASP..107..803U}). Their non-thermal emission is generally detected across the entire electromagnetic spectrum from radio to $\gamma$-ray bands. Blazars are sub-classified into flat spectrum radio quasars (FSRQs) and BL Lac objects (BL Lacs), according to the equivalent width of the emission lines in their optical spectrum (\citealt{1991ApJ...374..431S}; \citealt{1991ApJS...76..813S}; \citealt{1996MNRAS.281..425M}). These two subclasses of blazars are thought to be intrinsically different, perhaps based on their accretion mode \citep{2016CRPhy..17..594D}. FSRQs have high luminosity and a thin and radiatively efficient black hole accretion disk \citep{1986ApJ...300..216M}, while BL Lacs are powered by an advection-dominated, low-radiative-efficiency accretion flow (\citealt{2016CRPhy..17..594D}; \citealt{2019ARA&A..57..467B}). The jet emission is relativistically beamed \citep{2019RLSFN.tmp...28G}, with a Doppler boosting factor corresponding to a bulk Lorentz factor of several to greater than ten \citep{2009A&A...507L..33P}. 
In both cases, the broad-band spectra consist of two broad humps, one peaking in the IR-to-X-ray regime, and the other peaking in the $\gamma$-ray regime. The low energy peak is believed to be due to synchrotron emission, while the high energy peak is likely due to inverse Compton scattering of low-energy photons of either the same synchrotron photons (for BL Lacs) or external photons from the disk/BLR (for FSRQs) (e.g. \citealt{2017ApJ...835..182D}). However, some blazars might not necessarily be detected in $\gamma$-rays (e.g. \citealt{2017ApJ...851...33P}).  Indeed, a recent study showed that blazars undetected in $\gamma$-rays are likely to have relatively smaller Doppler factors and more disk dominance \citep{2017ApJ...851...33P}. In the case of strong Compton scattering, the beaming of $\gamma$-rays could be larger than e.g. that seen in  the radio \citep{1995ApJ...446L..63D} leading to the possible non-detection (or reduced detection efficiency) of $\gamma$-rays from sources not seen exactly pole-on.

The \Fermi Gamma-ray Space Telescope has been observing the high-energy sky from 50 MeV to 1 TeV since 2008 \citep{2009ApJ...697.1071A}. The fourth \Fermi LAT catalog (4FGL) of $\gamma$-ray sources is based on observations for eight years, and is a deep all-sky survey in $\gamma$-ray bands \citep{2020ApJS..247...33A}. The \Fermi 4FGL catalog contains 5064 $\gamma$-ray sources detected with at least 4$\sigma$ significance. More than 62\% of these sources are associated with AGN. Currently, 26\% of \Fermi 4FGL sources are unassociated.  The high-galactic latitude sources may fall in the following categories: blazars, radio galaxies, and milli-second pulsars (e.g. \citealt{2020ApJS..247...33A}; \citealt{2017ApJ...838..139S}), with the largest fraction of high-Galactic-latitude (and presumably extra-galactic) sources being blazars, where millimeter-wave (mm) and $\gamma$-ray emission are likely to be produced co-spatially in the extremely compact emission regions within the jet \citep{2019ApJ...877...39M}. This makes the mm regime a very efficient way to identify both blazars and previously unassociated \Fermi sources.

Previous studies have targeted monitoring of \Fermi $\gamma$-ray blazars at various wavelengths, based on, for example, follow-up at radio wavelengths (\citealt{2015ApJS..217....4S}; \citealt{2017ApJ...838..139S}) or the infrared behavior of $\gamma$-ray sources with \textit{WISE} (\citealt{2012ApJ...748...68D}; \citealt{2016ApJ...827...67M}). From these studies, blazars are known to reside at high redshifts ($z>0.1$) and exhibit extreme apparent luminosities with strong variability.

The South Pole Telescope (SPT, \citealt{2011PASP..123..568C}) is a 10-meter telescope dedicated to studying the cosmic microwave background (CMB). The SPT has been used to survey thousands of square degrees of the southern extragalactic sky at 1.4, 2.0, and 3mm with arcminute resolution down to milli-Jansky noise levels. In this work, we focus on the 2500 $\mathrm{deg^2}$ SPT-SZ survey, conducted with the SPT from 2008-2011. The SPT-SZ field is at high Galactic latitude ($|b|>15$), and thus, most of the sources are expected to be of extra-galactic origin \citep{2020arXiv200303431E}.  Roughly 3500 synchrotron-dominated sources are detected at high significance in the SPT-SZ maps, providing a powerful tool to identify unassociated 4FGL sources. This wavelength regime is particularly suited to uncovering the extragalactic unidentified \Fermi 4FGL sources, because the mm sources are almost exclusively ($>$80\%) blazars, and blazars are the dominant population in the $\gamma$-ray regime. The instrument sensitivity, survey area, and resolution of the SPT are well suited to this task.

In this paper, we establish the methodology of associating mm point sources from CMB surveys with $\gamma$-ray sources. In \autoref{sec:data} we present the data used in this study, while \autoref{sec:method} describes the association method. \autoref{sec:results_discussion} discusses the implications for the associations, the multi-wavelength properties of the new associations, and the comparisons to previous work. In this analysis, we use a flat $\mathrm{\Lambda CDM}$ cosmology with $\Omega_{\Lambda}=0.73$, $\Omega_{M}=0.27$, and $H_0=71\ \mathrm{km\ s^{-1}\ Mpc^{-1}}$.

\section{Observational Data}
\label{sec:data}

We use multi-wavelength data to associate and characterize the 282 4FGL sources within the 2500 $\mathrm{deg}^2$ SPT-SZ survey field (\autoref{fig:fields}). We note that the SPT-SZ field is at high Galactic latitude and thus it is assumed that the vast majority of the sources are of extra-galactic origin. The data sets used in our analysis are summarized below.

\subsection{Gamma rays}
We rely on the  $\gamma$-ray data from the \Fermi 4FGL catalog \citep{2020ApJS..247...33A}.   We use the source coordinates and 95\% uncertainty ellipse for source cross-matching and assume the beam to be Gaussian. The median effective radius for the sources within the 2500 ${\rm deg^2}$ SPT-SZ survey field is 3.2 arcmin.  The 0.1--100\,GeV energy flux and its uncertainty are used to study the multi-wavelength flux correlation and the spectral properties of the associated sources. The class designation in 4FGL is used to evaluate the multi-wavelength associations. 

\subsection{X-ray}
X-ray data are from the \textit{ROSAT} All-Sky Survey (RASS) Bright Source Catalog \citep{1999A&A...349..389V} and the Faint Source Catalog \citep{2000yCat.9029....0V} at X-ray energies 0.1-2.4 keV. The source coordinates and  count rate are both used for 4FGL association. We also use source count rate to analyze potential flux correlation. The statistical signal-to-noise ratio reported in the catalog is used when we plot the uncertainties of count rates.  Because of the high Galactic latitude of the SPT-SZ field ($|b|>15$), the soft X-ray \textit{ROSAT} data are not affected by photoelectric absorption from our own Galaxy.  In addition, blazars generally don't exhibit significant intrinsic soft X-ray absorption \citep{2005ApJ...625..727P}. Therefore, the \textit{ROSAT} X-ray measurement should be a reliable measure of the X-ray brightness of the blazars, although dependent on the absorption along the line of sight and the different components sampled in blazars of different classes (e.g. synchrotron emission in BL Lacs and inverse Compton emission in FSRQs). \lz{However, the X-ray brightness of a blazar may depend on the blazar sub-class and which component (synchrotron or inverse Compton scattering) is being sampled by \textit{ROSAT}.}

\subsection{Infrared}
Infrared data are taken from the Wide-field Infrared Survey Explorer (\textit{WISE}) AllWISE Source Catalog at 3.4, 4.6, 12, and 22 $\mathrm{\mu m}$ (\citealt{2010AJ....140.1868W}; \citealt{2013yCat.2328....0C}). The angular resolution of WISE ranges from 6\arcsec\ to 12\arcsec\ from short to long wavelengths. The source coordinates and flux at 22 $\mathrm{\mu m}$ (W4) are both used for 4FGL association. The flux at 22 $\mathrm{\mu m}$ is also used to study potential flux correlation. The 4-band magnitudes are used to perform the \textit{WISE} color analysis on 4FGL blazars. Both the statistical noise reported in the catalog and and additional 10\% uncertainty from calibration in W4 \citep{2010AJ....140.1868W} are included when we plot the  flux uncertainty.

\subsection{Millimeter}
Millimeter-wave point sources are from the 2500 $\mathrm{deg}^2$ SPT-SZ survey \citep{2020arXiv200303431E} which has a spatial resolution of 1\arcmin.15 at 150 GHz (2 mm) and an absolute astrometric uncertainty of 2\arcsec\ \citep{2010ApJ...719..763V}.  The source coordinates and flux at 150 GHz are both used for 4FGL association. The flux at 150 GHz is also used to study the flux correlation and spectral classification. The flux uncertainty includes the statistical noise reported in the catalog, 1.15\% uncertainty for the absolute calibration at 150 GHz and 20\% uncertainty from mm source variability. The 20\% variability is the median value of the variance in the light curves of the brightest 200 blazars in the SPT-SZ field. The SPT data used in this paper were collected between 2008 and 2011. Note that the published SPT-SZ catalog reaches down to the $4.5\sigma$ significance level, but in this work we extended our search down below $4.5\sigma$ to  $1\sigma$ using forced photometry directly from the 150 GHz maps using the SUMSS positions as priors. This combination of SUMSS (radio) and SPT (mm) associations will hereafter be referred to as SPT+SUMSS counterparts.

\subsection{Radio}
Radio data are from the Sydney University Molonglo Sky Survey (SUMSS, \citealt{2003MNRAS.342.1117M}) at 843 MHz with 45\arcsec\ angular resolution and a detection threshold of 6 mJy. The data were taken between 1997 and 2003. The source coordinates and integrated radio flux density are both used for 4FGL association. The flux density is also used to study the flux correlation and spectral classification. In addition to the statistical noise reported in the catalog, we also adopt a 20\% uncertainty to account for source variability. This uncertainty is what we observe for sources in the mm on roughly year-long time scales, but note that this is most likely an underestimate for the source variability on the time scales we are comparing fluxes for this work.

\subsection{Spectroscopic Redshifts}
For each associated 4FGL source, we gather all multi-wavelength data and adopt the most accurate position from radio/mm/infrared/optical counterpart if available. We obtained spectroscopic redshifts from the NASA/IPAC Extragalactic Database (NED) for any sources for which they were available. Using the archival redshift data, we study the luminosity distribution, multi-wavelength flux evolution, and classification distribution of the 4FGL blazars.

\section{Method}
\label{sec:method}

\begin{figure}
\centering
\includegraphics[width=0.45\textwidth]{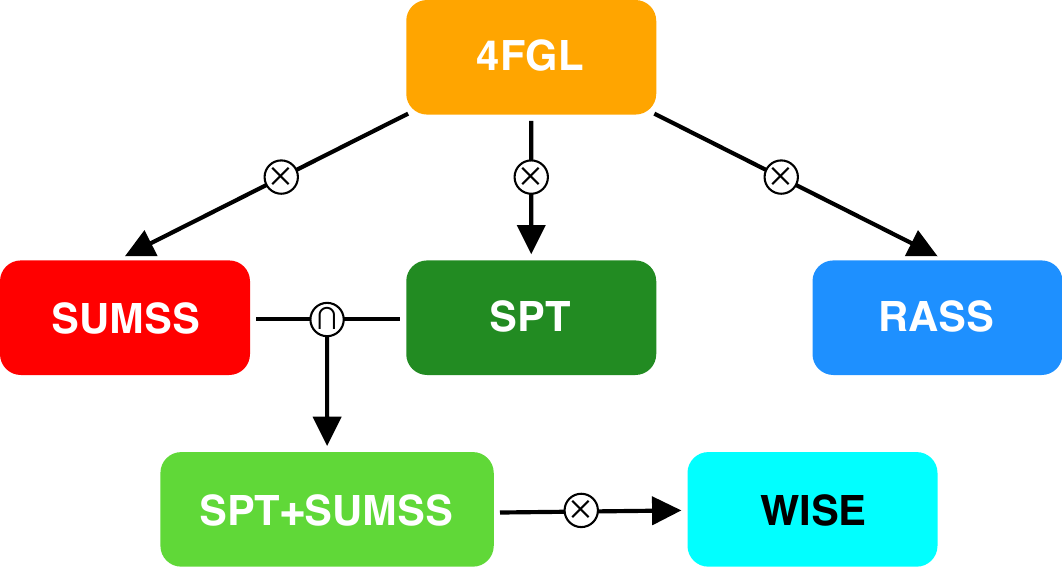}
\caption{The logic sequence of multi-wavelength cross-matching of 4FGL sources. We initially conduct the preliminary 4FGL cross-matching with SUMSS, SPT, and RASS to evaluate the most efficient wavelength for the $\gamma$-ray source identification, where the SPT+SUMSS association turns out to maximize the completeness while minimizing the impurity. The true counterparts are selected based on the statistics of the \Pvalue from the preliminary results. We then take the advantage of mm-radio (SPT+SUMSS) counterparts to refine the position and enable a reliable cross-match with WISE. }
\label{fig:diagram}
\end{figure}

\begin{figure*}
\begin{center}
    \includegraphics[width=1.00\textwidth]{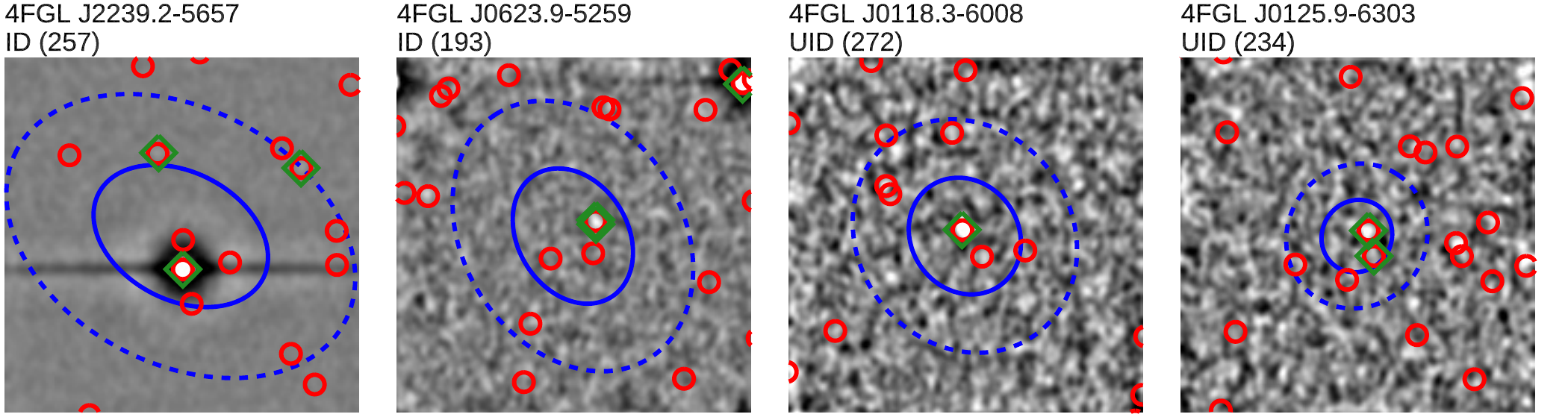}
\end{center}
\caption{Selected samples illustrating various kinds of 4FGL associations with mm and radio sources. 4FGL source name and thumbnail index (see Appendix for all the sources studied in this work) are shown at the top-left of each panel. Each $0.7^{\circ} \times 0.7^{\circ}$ thumbnail in the grey background is the high-pass filtered SPT 150 GHz image.
The \textit{blue ellipse} at the center show the 4FGL 95\% uncertainty position and the dashed blue ellipse represents the $4\sigma$ uncertainty. The \textit{green diamond} marks the position of the SPT point source. The \textit{red circle} shows the position of SUMSS point sources. The first two examples are of previously associated 4FGL sources, which already had known counterparts. In both cases, these known sources are also uniquely identified by SPT (the first is very bright in the mm waveband, while the second is fainter).  The last two panels show previously unassociated 4FGL sources. The third panel shows a new unique association with an SPT source. The last panel shows an unusual case where two SPT sources fall within the 4FGL error circle, either or both of which could be producing the $\gamma$-ray emission. 
}
\label{fig:ex_method}
\end{figure*}

\begin{figure*}
\begin{center}
    \includegraphics[width=0.95\textwidth]{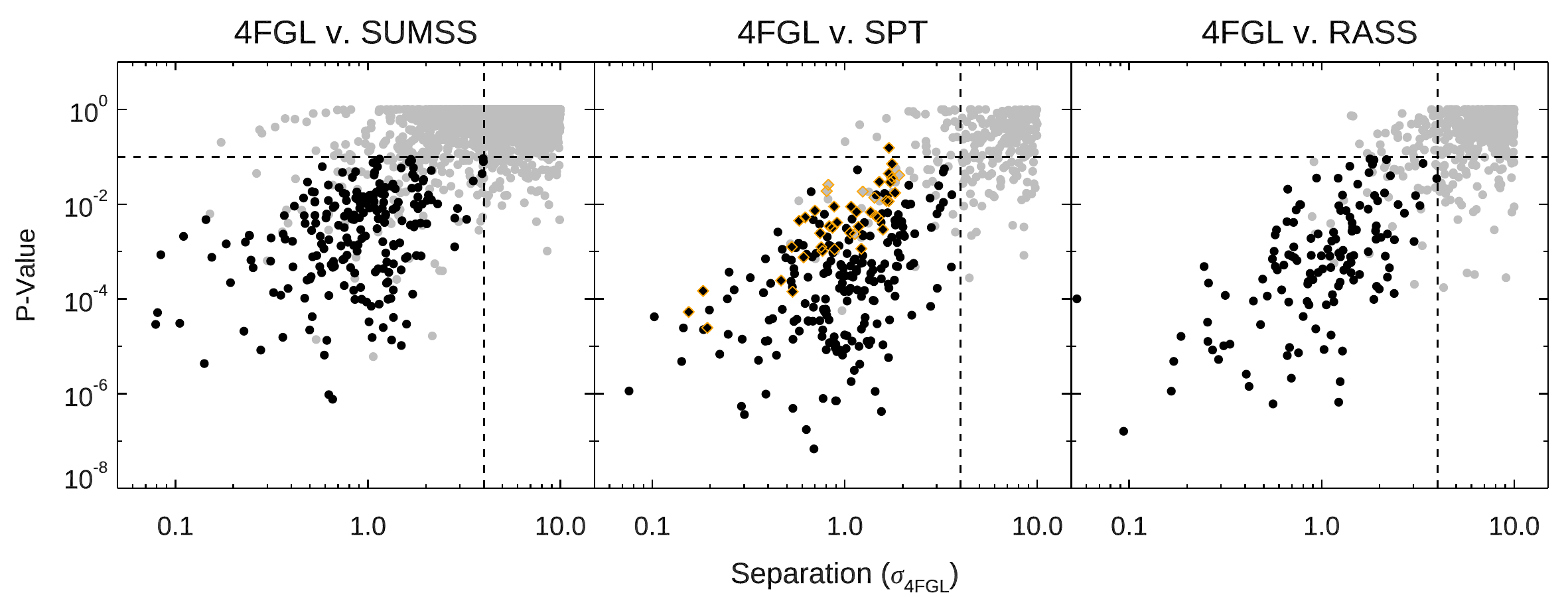}
\end{center}
\caption{Potential associations of the sources from external catalogs with the 282 4FGL sources in the SPT-SZ survey field plotted with their probability of false association (\pvalue) versus source separation in units of positional uncertainty. The \textit{grey} dots represent all the point sources within $10\sigma$ 4FGL positional uncertainty. The vertical and horizontal dashed lines indicate our adopted $r<4\sigma_{\mathrm{4FGL}}$ positional separation cut and \pvalue$<0.1$ cut, respectively. The \textit{black} dots represent the source with the lowest \pvalue associated to a 4FGL source. \textit{orange} diamonds represent faint mm sources with fluxes assigned by forced photometry using positional priors from the SUMSS radio catalog. Note that $\sigma_{\mathrm{4FGL}}$ refers to the standard deviation of the 4FGL position, which varies source-by-source. 
\textbf{Left Panel}: The SUMSS radio associations are heavily contaminated due to the excessively high source number density; thus, the region of spurious association extends from the top-right region to roughly $4\sigma_{\mathrm{4FGL}}$ in separation and 0.1 in probability of false association. 
\textbf{Middle Panel}: The SPT mm association shows more distinct separation between the two groups of the associations in $p$-$r$ space, while for radio association, the contamination of the spurious cross-matching causes the large overlap in the spurious region. With both the high completeness and purity, mm provides the best combined performance in completeness and reliability.
\textbf{Right Panel}: The RASS X-ray association shows good purity according to the low probability of false association (most grey dots within $4\sigma_{\mathrm{4FGL}}$ have \pvalue less than 0.1) but the completeness is the lowest compared with the others (see \autoref{tab:completeness}). Multi-wavelength comparisons of sources within $4\sigma_{\mathrm{4FGL}}$ in single parameter space are histogrammed in \autoref{fig:dr_p_histogram}. Note that the sources shown at low \pvalue and separation $<4\sigma_{\mathrm{4FGL}}$ which are shown as grey are where there are multiple counterparts for a given 4FGL source and we have adopted the counterpart with the lowest \pvalue as the association. }
\label{fig:dr_p}
\end{figure*}

\begin{figure*}
\begin{center}
    \includegraphics[width=1.00\textwidth]{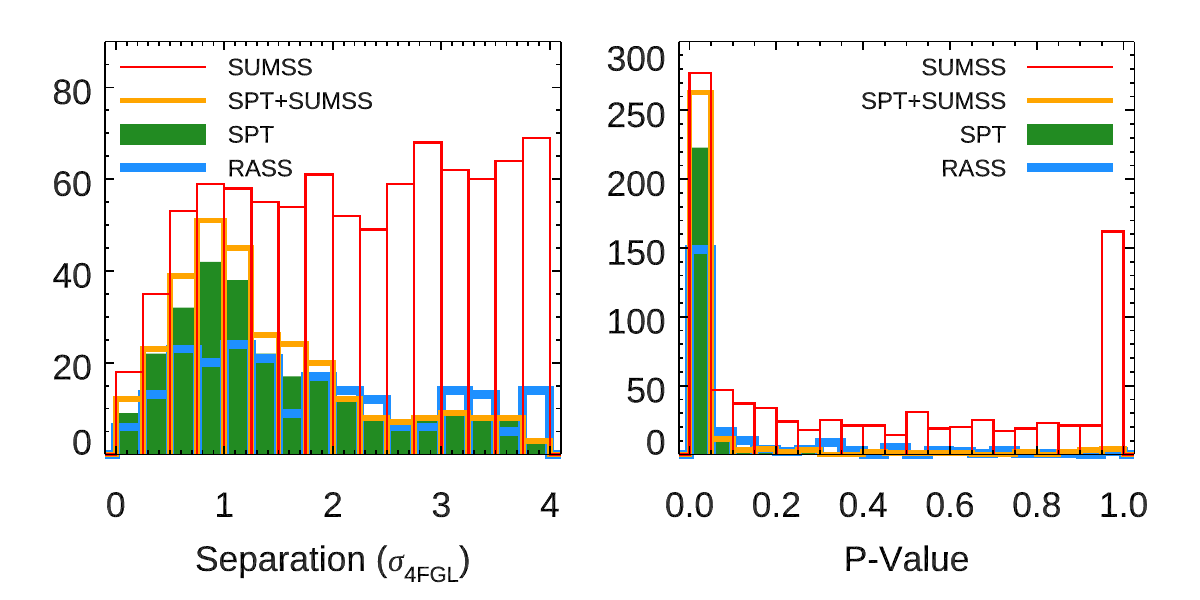}
\end{center}
\caption{Multi-wavelength comparison of the potential associations with 282 4FGL sources in the SPT-SZ survey field. Y-axis for both panels are number of sources. These histograms indicate that SPT+SUMSS is the most efficient identifier of $\gamma$-ray counterparts with high completeness and reliability. \textbf{Left Panel}: Histogram of angular separation of 4FGL sources from SUMSS (radio), SPT (mm), SPT+SUMSS (mm), and RASS (X-ray) positions. mm and radio associations are highly complete, while X-ray only has around half of their completeness. 
\textbf{Right Panel}: Histogram of the probability of false associations of the 4FGL sources with SUMSS, SPT, SPT+SUMSS, and RASS sources. A low probability of false association corresponds to a high certainty of a real counterpart. Thus, the mm and X-ray counterparts are more secure because most of them have the probability of false association less than 0.1, while only around $2/3$ radio potential counterparts are within 0.1.}
\label{fig:dr_p_histogram}
\end{figure*}

In this section, we describe the method we use for identifying and associating multi-wavelength counterparts to the \Fermi 4FGL sources. There are 282 4FGL sources within the $2500~\mathrm{deg}^2$ SPT-SZ survey field, 71 (25\%) of which are previously unassociated with any counterpart.  Attempting to cross-match the 4FGL sources to external catalogs by selecting all potential counterparts within a given search radius will often yield multiple potential counterparts, particularly when the external catalog has a high surface number density.  There are two quantities associated with every external source that we use for cross-matching. The first is the separation between the potential counterpart and the 4FGL position in units of positional uncertainty ($\sigma_{\rm 4FGL}$). The second is the  probability of false association for a given external source, which is a function of separation, flux density, and catalog density.

In \autoref{fig:diagram}, we show the logical flow diagram we use to identify the multi-wavelength counterparts, which is described in more detail later in this section. 
\autoref{fig:ex_method} shows the 4FGL positional uncertainty ellipses on top of representative thumbnail images from the SPT-SZ survey and also the positions of radio sources from the SUMSS catalog. (see \autoref{sec:thumbnails} for the thumbnails of the entire sample.)

\subsection{Probability of False Association}

To start our cross-matching analysis, we search for external candidates for each source in the 4FGL catalog within its $4\sigma$ positional error ellipse. The choice of $4\sigma$ was a qualitative choice, designed to be inclusive of the relatively large positional uncertainties of both \Fermi and the external catalogs.

In order to robustly identify the most probable counterpart when there are multiple candidates, and to statistically evaluate the odds of spurious associations, we calculate a simple \Pvalue (\citealt{1978MNRAS.182..181B}; \citealt{1986MNRAS.218...31D}; \citealt{2011MNRAS.413.2314B}), which gives the probability of a matched source being randomly associated within a given area. The \Pvalue takes into account the angular separation, flux densities, and the number density of potential counterparts. The expected number of random associations is:
\begin{equation}
	\mu = \pi a b n_{s}(>S)
	\quad, 
\end{equation}
where $a$ and $b$ are semi-major and semi-minor axes, respectively, and $n_{s}(>S)$ is the surface number density of sources brighter than the candidate counterpart. The \Pvalue (hereafter ``\pvalue''), is thus defined as:
\begin{equation} 
    p = 1-e^{-\mu}
    \quad.
\end{equation}
The \pvalue constructed in this way is the probability that an association is false, due to spurious coincidence. Thus, if $p \ll 1$, the association is highly likely to be real. We note that there are small corrections that can be made to this probability (see, e.g. \citealt{1986MNRAS.218...31D,2011MNRAS.413.2314B}), which we neglect in this work as the density of sources is relatively low and our sources of interest are rare.  In this work, we accept all sources as potential counterparts with \pvalue$<0.1$ within the $4\sigma$ association area. 

To visualize our cross-matching selection procedure, \autoref{fig:dr_p} shows all the sources within $10\sigma$ of the 4FGL position with the $4\sigma$ 4FGL and \pvalue $< 0.1$ cuts indicated by dashed lines. We plot the \pvalue versus the angular separation in units of positional error ($\sigma_{\rm 4FGL}$) for each of the external catalogs.  In some cases there are multiple counterparts that meet this criteria. When this is the case, we adopt the counterpart with the lowest \pvalue as the most probable counterpart and show that source in  \autoref{fig:dr_p} with a black point.

\subsection{Completeness and Purity of Individual Catalogs}

\begin{table*}
\centering
\caption[SPT-SZ Field]{Completeness and purity of 4FGL multi-wavelength associations within the 2500 $\mathrm{deg}^2$ SPT-SZ survey sky} 
\renewcommand{\arraystretch}{1.5}
\begin{tabular}{ l c c c c c c c c }
\hline\hline
\multicolumn{2}{l}{\thead{Survey\\\ }} & \thead{$\Sigma$\\($\mathrm{deg}^{-2}$)} & \thead{4FGL (282)\\\ } & \thead{4FGL-ID (211)\\\ } 	& \thead{4FGL-UID (71)\\\ }  & \thead{Completeness\\(\%)}  & \thead{Purity\\(\%)} \\
\hline
\multicolumn{2}{l}{RASS}  & 3.53  & 136 & 119 & 17 & 48.2$\pm$4.1 & 89.8$\pm$0.23 \\
\multicolumn{2}{l}{SPT}   & 1.91  & 204 & 181 & 23 & 72.3$\pm$5.1 & 93.7$\pm$0.01 \\
\multicolumn{2}{l}{SPT+SUMSS}   & 1.94  & 239 & 199 & 40 & 84.8$\pm$5.5 & 94.4$\pm$0.02 \\
\multicolumn{2}{l}{SUMSS} & 26.75 & 232 & 198 & 34 & 82.3$\pm$5.4 & 58.9$\pm$3.06 \\
\hline
\end{tabular}
\label{tab:completeness}
\tablecomments{Three wavelengths -  X-ray (RASS), mm (SPT and SPT+SUMSS), radio (SUMSS) - are used to study the performance of the 4FGL association within the 2500 $\mathrm{deg}^2$ sky covered by the SPT-SZ survey. The second column refers to the surface number density of each survey. The third to fifth column represent the total number of potential counterparts for all 4FGL sources, previously identified 4FGL sources, and previously unassociated 4FGL sources, respectively. The last two columns are completeness and purity for the multi-wavelength association, where the error of completeness is Poissonian and error of purity is evaluated from bootstrapping. }
\end{table*}

In \autoref{fig:dr_p_histogram}, we quantify the behavior of the multi-wavelength associations in the space of angular separation and the \pvalue by histograming potential multi-wavelength counterparts of 4FGL point sources within their $4\sigma$-beam \textit{before} cutting on \pvalue. 

In the left panel of \autoref{fig:dr_p_histogram}, we show the histogram of angular separation in units of positional error for each of the external catalogs within the adopted cutoff of $4\sigma_{\rm 4FGL}$. What is immediately apparent is that the curves are mostly flat above $1\sigma_{\rm 4FGL}$ for SUMSS and RASS, but peaks around $1\sigma_{\rm 4FGL}$ for SPT (or SPT+SUMSS), as the separations increase. In the case of SUMSS, the sources density is high ($\sim30$ sources per square degree) and so there are always potential counterparts to match with, and there are more the further out you include, and so the problem becomes in deciding which is the true counterpart. For RASS, there are simply not many counterparts to match with. SPT (or SPT+SUMSS), however, often has a counterpart to 4FGL, and the density of background sources is lower, so the association histogram peaks and then decreases at larger separations.

In the right panel of \autoref{fig:dr_p_histogram}, we plot the histogram of \pvalue for all sources within the 4$\sigma_{\rm 4FGL}$ radius. The distribution of SUMSS counterparts is bimodal, with the counterparts split into two groups near 0 and 1 respectively, which indicates that nearly $1/3$ of SUMSS sources inside $4\sigma$ 4FGL beams are likely to be spurious associations. The SPT (or SPT+SUMSS) and RASS catalogs, however, with much lower densities, have just one peak at low \pvalue, indicating a high certainty of association.

We adopt the critical \pvalue ($p_{\mathrm{cri}}$) to best separate the real and false associations. We chose 10\% as an acceptable false association rate and indicate this cut by the horizontal dashed lines in \autoref{fig:dr_p}.

For the purposes of defining the efficiency of an external catalog associating with 4FGL sources, we define completeness as the probability of a catalog providing at least one viable counterpart (separation $<4\sigma_{\rm 4FGL}$ and \pvalue$>0.1$) to a source in the 4FGL catalog, and the purity as the probability that the association is false (separation $<4\sigma_{\rm 4FGL}$). To calculate the purity of each catalog we use the equation: 
\begin{equation}
    {\rm Purity} = 1 - \frac{1}{N}\sum_{r<4\sigma_{\rm 4FGL}}p
    \quad,
\end{equation}
where $N$ and $r$ refer to number of samples and angular separation, respectively. 

The X-ray (RASS) associations have a high purity (\autoref{fig:dr_p_histogram} right panel) but only 54\% 4FGL sources have RASS sources inside their $4\sigma$ uncertainty regions, and is thus largely incomplete.  

Matching the 4FGL sources to the SUMSS catalog we find that 95\% of the 4FGL sources have at least one SUMSS source falling inside their 4$\sigma$ uncertainty region.  However, as \autoref{fig:dr_p} demonstrates, many SUMSS sources which are within the $4\sigma$ 4FGL uncertainty regions also have a high probability of being spurious associations, i.e. they lie near $p=1$ and $r<4\sigma_{\rm 4FGL}$ in $p$--$r$ space. \autoref{fig:dr_p_histogram} also demonstrates that $\sim1/3$ of SUMSS associations have high probability of false association (\pvalue$>0.1$) and there are often multiple possible radio counterparts within the 4$\sigma$ uncertainty region. We find that $\sim$55\% of 4FGL sources have more than one potential SUMSS counterpart within the 4FGL positional uncertainty. This demonstrates that while the radio catalog has a high completeness, it also has a low purity (i.e. low confidence of a true association).  When we cut the sources within $4\sigma_{\rm 4FGL}$, we are left with 232 associations and a completeness of 82\%.

Matching the 4FGL sources to the SPT-SZ catalog we find that 204 (72\%) of the 4FGL sources have at least one SPT source falling inside their $4\sigma$ uncertainty regions and \pvalue$<0.1$.  Each SPT source was also cross-matched with SUMSS, within 1 arcmin of the SPT position at 150 GHz and with \pvalue less than 0.1. Nearly all (99\%) of the SPT sources that are located inside $4\sigma$ uncertainty regions of 4FGL have SUMSS counterparts. Only five 4FGL sources (\#80, \#127, \#177, \#211, and \#225 in \autoref{sec:long_table}. For details see thumbnails in \autoref{sec:thumbnails}) have SPT counterparts but no SUMSS counterparts. When we include the faint SPT fluxes at $>1\sigma$ derived from the SUMSS positions (SPT+SUMSS) we find an additional 35 have mm counterparts within the $4\sigma$ 4FGL beams and \pvalue$<0.1$, bringing the completeness to 85\%.  The mm associations also have high completeness, but because of the far lower surface density of SPT sources, the associations retain a high purity (see \autoref{fig:dr_p_histogram}, right panel).  Among the 71 previously unassociated 4FGL sources in this region, 40 (56\%) have mm counterparts in the SPT-SZ data.

In \autoref{tab:completeness}, we show the completeness and purity of each of the catalogs we associate to the 4FGL sources after the \pvalue cut. We estimate the purity for each association as discussed earlier.  The completeness is defined as the fraction of 4FGL sources with a probable counterpart. In the results of multi-wavelength associations, the mm (including the faint sources from the forced photometry using SUMSS priors) and radio bands have high completeness of association (85 \% and 82 \%, respectively), while the X-ray band is less than half complete (48 \%).

We thus conclude that the SPT+SUMSS is the most efficient means to identify 4FGL sources, as it maximizes both completeness and purity. Almost all of the SPT identified sources (99\%) have corresponding SUMSS counterparts which appear to be flat-spectrum at mm and radio wavelengths (\autoref{fig:s843_s150}). 
 
The results of multi-wavelength associations demonstrate the promise of mm association to identify $\gamma$-ray sources. Thus, the combination of SUMSS (radio) and SPT (mm) associations (SPT+SUMSS) can best characterize the 4FGL sources because of the high certainty and high completeness of $\gamma$-ray association with joint mm and radio counterparts. For the previously identified 4FGL sources without SPT+SUMSS counterparts, we find that three of them are pulsars, one is a faint BL Lac, and the rest are undetermined blazar candidates with X-ray emission.

\subsection{The Construction of Our Combined 4FGL Multi-Wavelength Catalog}
\label{sec:multiwavelength_4fglcatalog}

To summarize the previous sections, we adopt the following criteria to select multi-wavelength counterparts of 4FGL sources for the analyses in this work (see \autoref{fig:diagram}): 
We select SUMSS/SPT(+SUMSS)/RASS counterparts within 4$\sigma_{\rm 4FGL}$ and \pvalue $<0.1$ as the 4FGL associated counterparts. 

\begin{figure}
\begin{center}
    \includegraphics[width=0.495\textwidth]{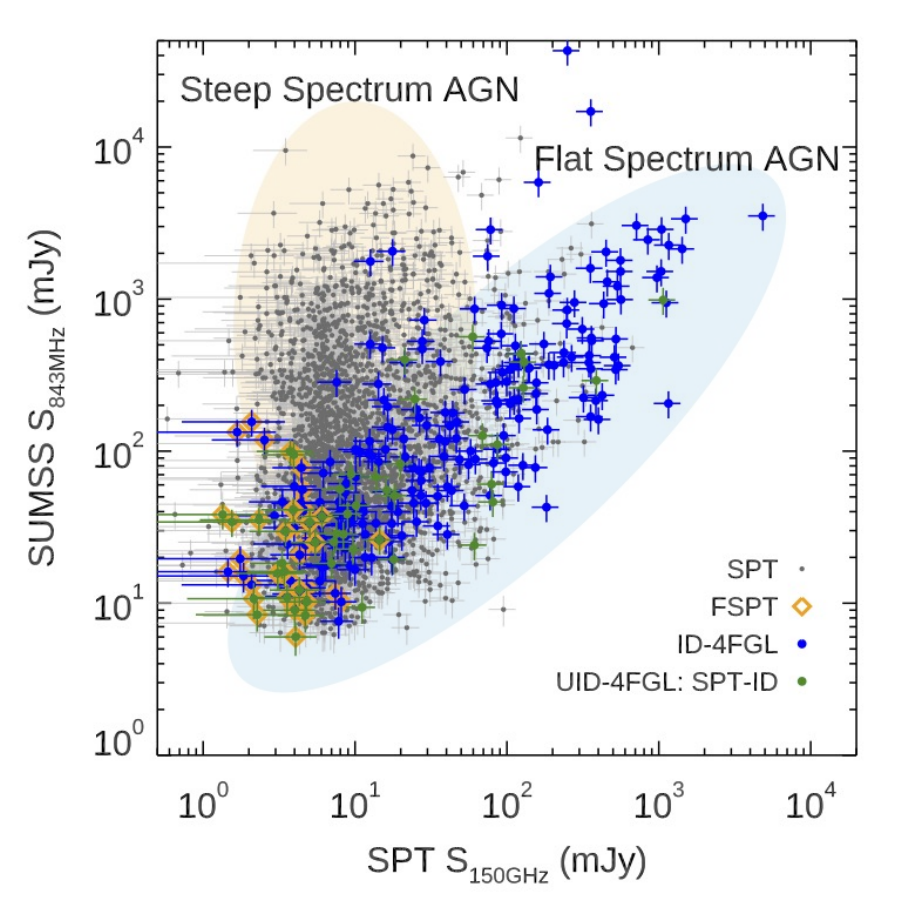}
\end{center}
\caption{The flux-flux plot showing the radio flux density at 843 MHz from the SUMSS survey versus the SPT flux density at 2 mm (150 GHz) for all SPT-selected sources. The \textit{grey} dots represent all the mm sources with radio counterparts. The \textit{blue/green} dots represent those with previous known/unknown $\gamma$-ray detection. The \textit{orange} diamonds represent faint mm sources with fluxes assigned by forced photometry using positional priors from the SUMSS radio catalog. Note that the steep-spectrum AGN roughly lie within the yellow oval and run up the vertical axis clustered around $S_{150\mathrm{GHz}}$=10 mJy, while the flat spectrum AGN lie within the blue oval which is nearly diagonal with slope=1. 239 4FGL sources are selected if they have both SPT and SUMSS counterparts. These sources are shown in green and blue. Note the total number of both blue and green dots in the plot is 265. This is because some 4FGL sources have multiple mm-radio counterparts as illustrated in \autoref{fig:ex_method}. Similar to \autoref{fig:fields}, the blue points are previously associated 4FGL sources, while the green points are previously unassociated 4FGL sources that have SPT counterparts. The majority of 4FGL sources are flat spectrum AGN. }
\label{fig:s843_s150}
\end{figure}

\begin{figure*}[ht]
\begin{center}
    \includegraphics[width=0.775\textwidth]{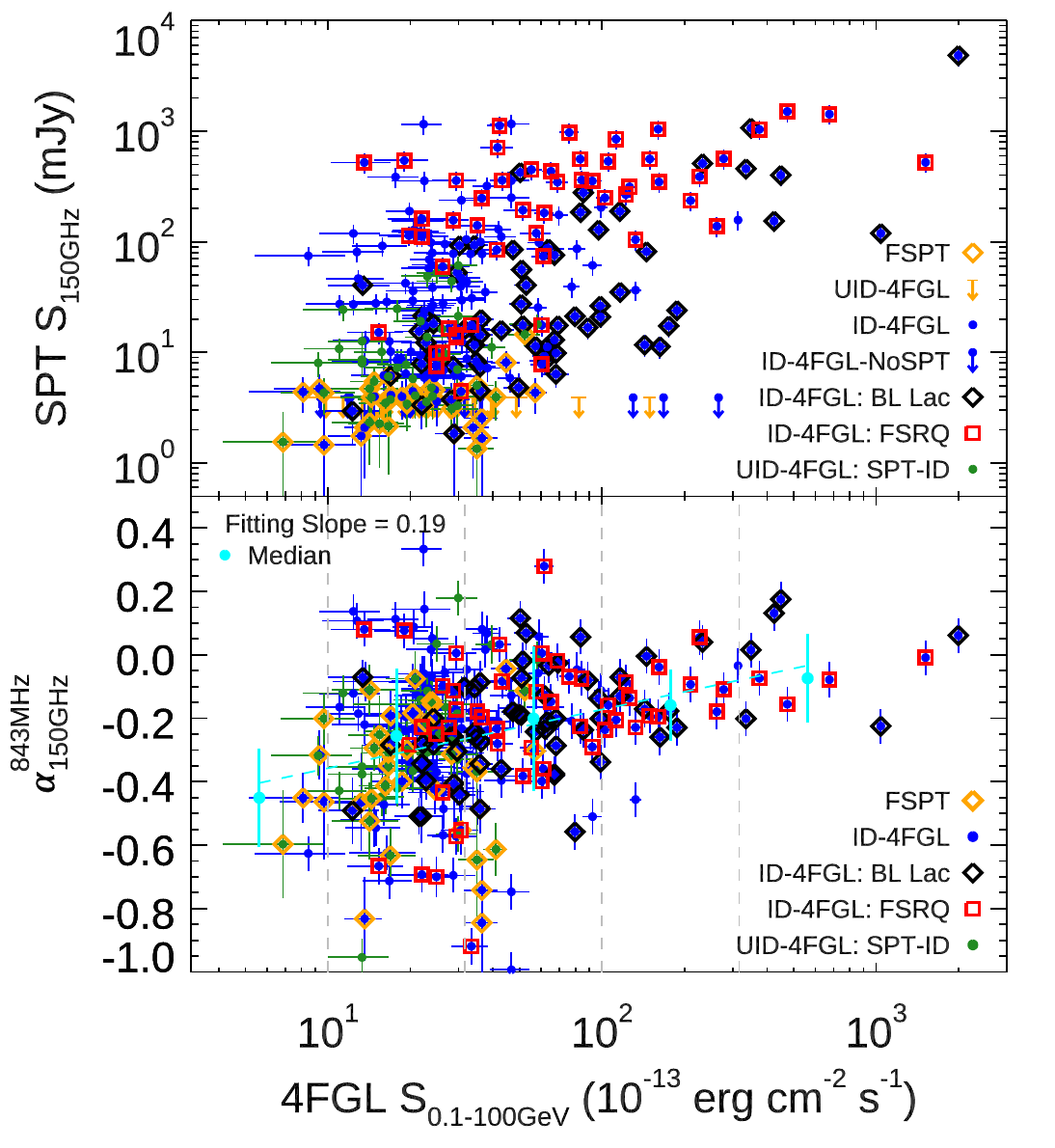}
\end{center}
\caption{
\textbf{Upper Panel}: mm and $\gamma$-ray flux correlation with each source labeled. Source types are labeled by their identities (previously identified 4FGL source: \textit{blue dot}; BL Lac: \textit{black diamond}; FSRQ: \textit{red square}; unassociated 4FGL source: \textit{green dot}; faint mm sources: \textit{orange diamonds}). 
The previously identified 4FGL sources without mm detections (ID-4FGL-NoSPT) and the remaining unidentified 4FGL sources (UID-4FGL) are indicated by \textit{blue} and \textit{orange} arrows, respectively, as $3\sigma$ upper limits from SPT-SZ. Note that three bright ID-4FGL-NoSPT sources ($S_{\rm 0.1-100GeV}>10^{-11}\ \mathrm{erg~cm^{-2}~s^{-1}}$) in blue arrows are previously identified pulsars and therefore fails to be identified by SPT. 
 Within associated 4FGL sources, FSRQs are more likely to be brighter than BL Lacs in mm wavelength. 
Most unassociated 4FGL sources are faint in both mm and $\gamma$-ray bands, which indicates that source identification might be limited by the sensitivity of the current generation of surveys. 
\textbf{Lower Panel}: The mm-radio spectral index ($\alpha_{\rm150GHz}^{\rm843MHz}$) as the function of $\gamma$-ray flux. Sources are labeled the same way as the left panel. \textit{Cyan dots} denote the median and standard deviations of the spectral index in each logarithmic bin of $\gamma$-ray flux. As shown in the plot, sources with brighter $\gamma$-ray fluxes are more likely to have flatter spectra except a few sources with $\gamma$-ray flux around $10^{-11}\ \mathrm{erg\ cm^{-2}\ s^{-1}}$. These outliers have excess radio emission compared with normal flat-spectrum AGN. Among these sources, some might have extra radio emission from nearby radio lobes; some are just ambiguous cross-matching that blends several radio counterparts; some have valid multi-band counterparts and need further investigation.}
\label{fig:flux_alpha}
\end{figure*}

The 4FGL-SUMSS cross-matching, while the most complete, is often degenerate with multiple possible associations.  
Recall that to help break the degeneracy of multiple possible SUMSS counterparts where there is no SPT source at $\ge 4.5\sigma$, we perform forced photometry at the SUMSS position in the SPT 150 GHz map. In this way, we are able to dig deeper into the SPT map and associate the most probable radio counterpart based on the mm flux. When there are multiple sources which meet the criteria of $4\sigma$-beams and \pvalue $<0.1$, we adopt the source with the lowest \pvalue in the mm. Some examples are shown in \autoref{fig:ex_method} to illustrate how the 4FGL sources are associated by SPT and SUMSS. Of the 239 4FGL-SPT matches, 204 sources are detected at $>4.5\sigma$ in \citet{2020arXiv200303431E}, while the remaining 35 have had mm-fluxes assigned using forced photometry using the SUMSS position as a prior. For clarity, these sources are highlighted in \autoref{fig:s843_s150} and \autoref{fig:flux_alpha} labelled as FSPT for ``faint".

Now, with this catalog of multi-wavelength associations in hand, we can associate the 4FGL sources to the WISE catalog to characterize their infrared emission.  Naive cross-matching 4FGL with \textit{WISE} is difficult because the high source density ($\sim$10$^4~\mathrm{deg}^{-2}$) results in high contamination from spurious associations. Each 4FGL source on average has $\sim$200 potential \textit{WISE} counterparts within the 95\% ($2\sigma$) uncertainty ellipse. However, SPT-selected blazars with a SUMSS counterpart can provide positional accuracy better than 15\arcsec \citep{2010ApJ...719..763V} when the signal-to-noise ratio is greater than 5 \citep{2007MNRAS.380..199I}.  Using the 4FGL-SPT-SUMSS associations, we performed the 4FGL association with \textit{WISE} based on the refined position, and selected the most likely candidates according to their lowest \pvalue. When cross-matching with \textit{WISE}, the \pvalue is especially useful to eliminate spurious false associations. We adopt \textit{WISE} associations using \pvalue$<0.03$ and separation within 9 arcsec of the combined SPT+SUMSS position. The characterization of the infrared emission of the 4FGL sources is discussed in Section \ref{sec:fermi_wise}.

Nearly all the SPT counterparts (99\%) of 4FGL sources also have a SUMSS counterpart allowing us to characterize each 4FGL source with both mm and radio fluxes. In the \autoref{fig:s843_s150}, mm sources with radio counterparts (grey dots) are roughly separated into two classes -- steep-spectrum AGN (yellow oval) and flat-spectrum AGN (blue oval). The majority of $\gamma$-ray sources (blue and green dots) have comparable mm and radio emission, which indicates that they are either BL Lacertae objects (BL Lacs) or flat-spectrum radio quasars (FSRQs).

\begin{figure}[ht]
\includegraphics[width=0.44\textwidth]{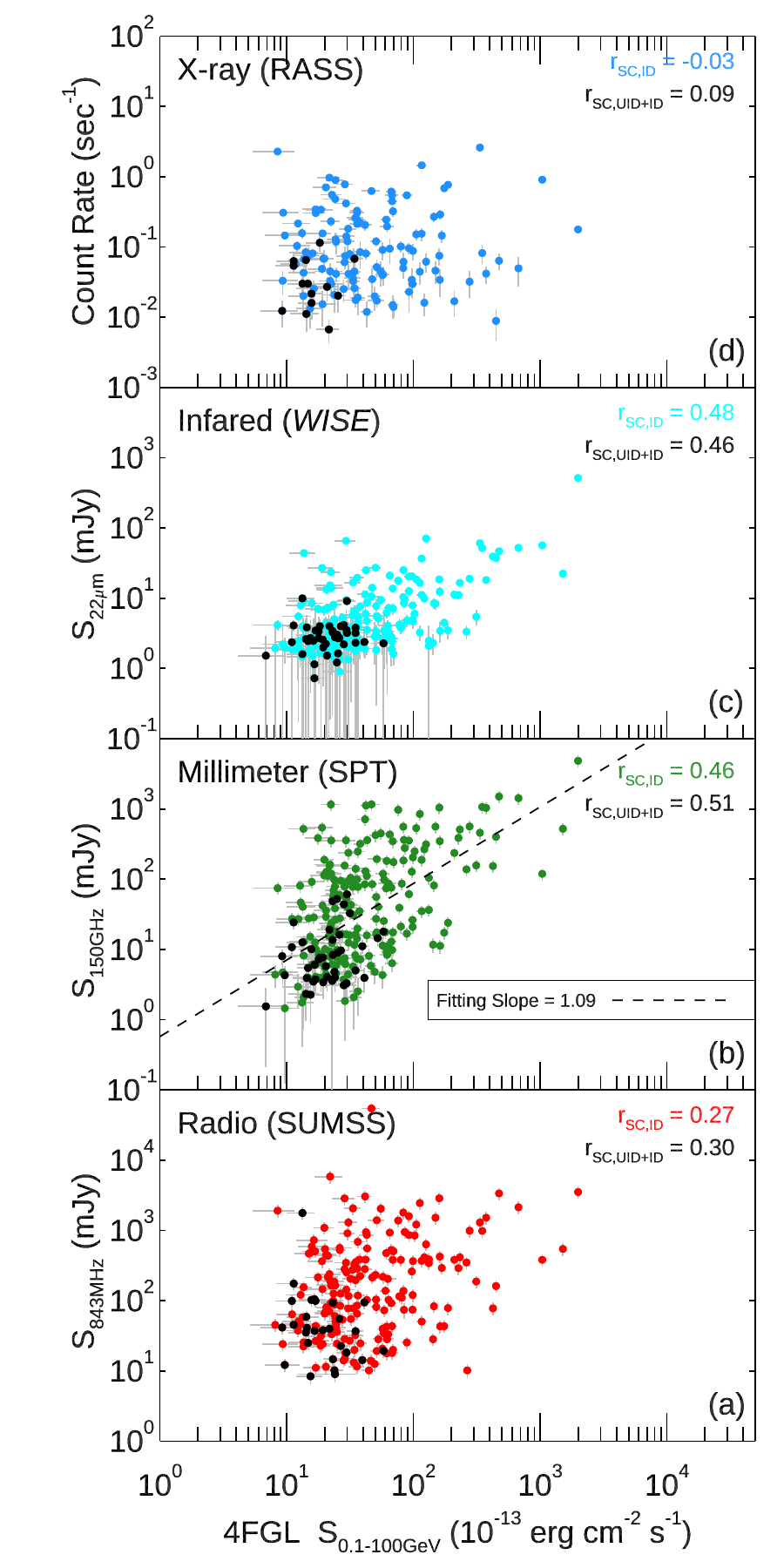}
\caption{Flux correlations between 4FGL and other surveys (X-ray: \textit{blue}; Infrared: \textit{light blue}; mm: \textit{green}; radio: \textit{red}), where \textit{black} dots indicate previously unassociated 4FGL sources with mm counterparts. The Spearman's rank correlation coefficient ($r_{\mathrm{SC}}$) of the flux correlation is shown at the top-right of each plot. Note that $r_{\mathrm{SC,ID}}$ refers to previously associated 4FGL sources and $r_{\mathrm{SC,UID+ID}}$ counts both previously associated and extra mm-identified 4FGL sources.  \textbf{Panel (a)}: As most of the 4FGL-SPT blazars are flat spectrum AGN, the $r_{\mathrm{SC}}$ in the radio-band should be similar that of the mm-band. However, $r_{\mathrm{SC}}$ in the radio-band is significantly lower than the mm-band because some 4FGL-SPT blazars have excess radio emission compared with most flat spectrum sources (see \autoref{fig:s843_s150}). 
\textbf{Panel (b)}: the mm and $\gamma$-ray bands are highly correlated, which is consistent with previous studies of multi-wavelength associations of 4FGL sources. The dashed line represents least squares fitting between the fluxes in the two bands. 
\textbf{Panel (c)}: The infrared-band has comparable correlation to the $\gamma$-ray band as the mm-band.  However, the association between 4FGL and \textit{WISE} has been refined by first matching to SPT because normal cross-matching would be heavily contaminated by the spurious associations due to the high source number density of \textit{WISE} ($\sim 10^4 \mathrm{deg}^{-2}$). 
\textbf{Panel (d)}: Unlike any other wavelengths, the X-ray fluxes are uncorrelated with $\gamma$-ray fluxes. This may be because the $\gamma$-ray traces the jet while the X-ray traces the coronal region of the AGN \citep{2017MNRAS.469..255G}.}
\label{fig:s_to_s}
\end{figure}

\begin{figure}
\begin{center}
    \includegraphics[width=0.475\textwidth]{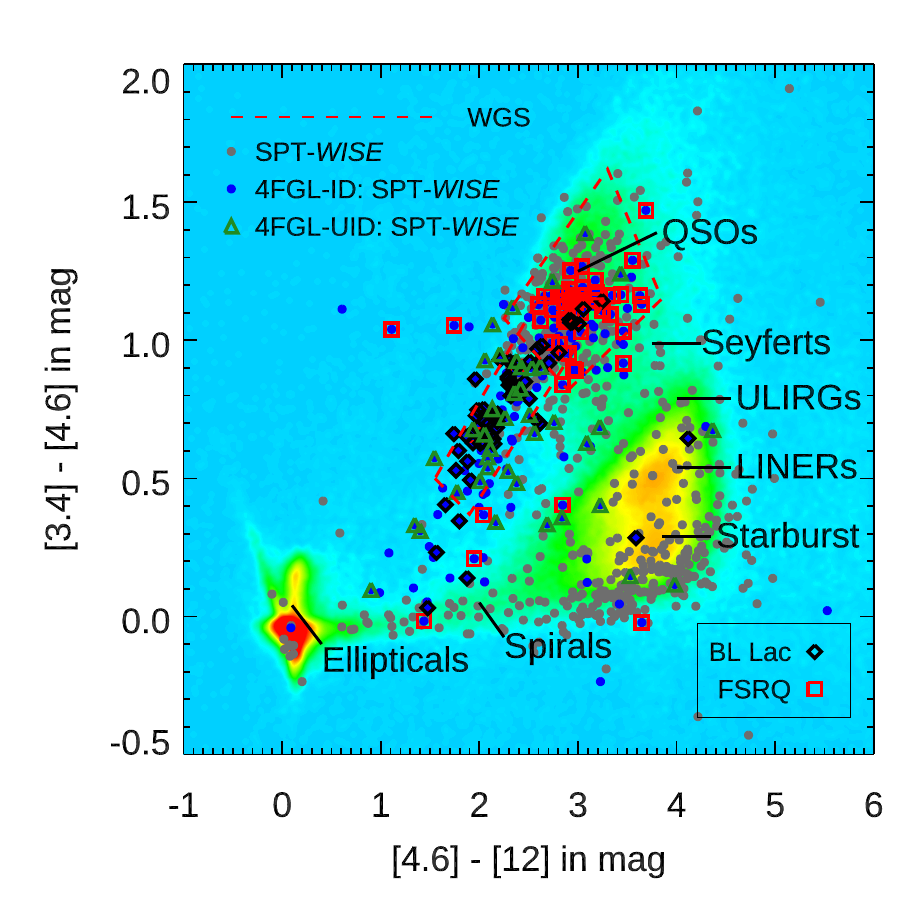}
\end{center}
\caption{\textit{WISE} color analysis. Background color map is based on the number source density in \textit{WISE} color space. This color map contours the classes of \textit{WISE} objects (see also \citealt{2010AJ....140.1868W}). On the background color map, each dot marks the SPT-selected source. These sources are also labeled by their identifications (previously associated 4FGL source: \textit{blue dot}; previously unassociated 4FGL source with mm counterpart: \textit{green triangle}; BL Lac: \textit{black diamond}; FSRQ: \textit{red square}). Red dashed lines outline the \textit{WISE} Gamma-ray Strip (WGS, \citealt{2012ApJ...750..138M}). Most of the mm associations lie inside the WGS, which indicates the consistency of mm correlation and WGS parameterization. Besides, the mm cross-matching provides more intrinsic associations and pinpoints the valid outliers outside WGS in \textit{WISE} color space.}
\label{fig:wise_color}
\end{figure}

\begin{figure}
\begin{center}
    \includegraphics[width=0.475\textwidth]{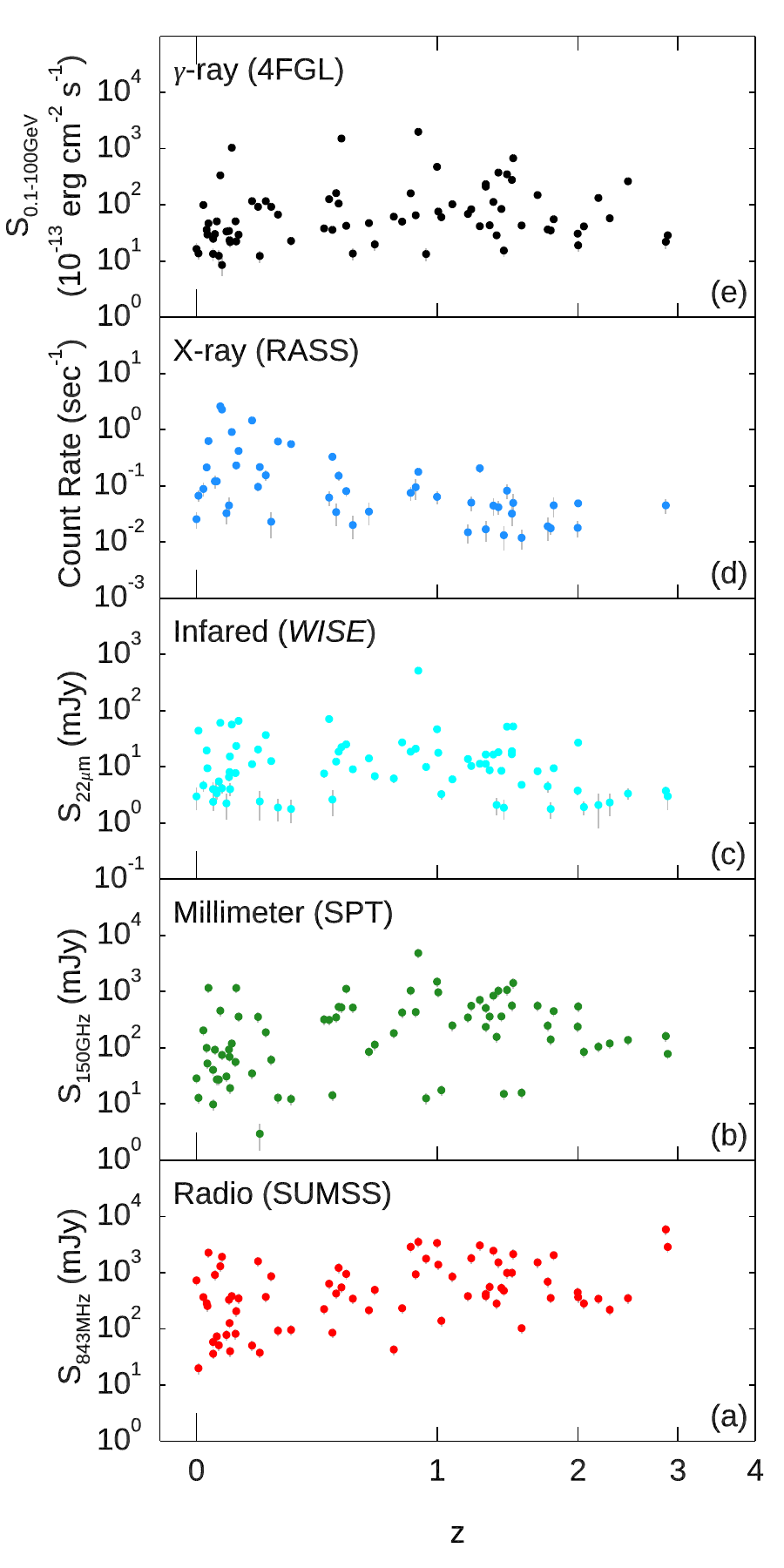}
\end{center}
\caption{Multi-wavelength flux density as the function of redshift. Most wavelengths (radio, mm, infrared, and $\gamma$-ray) show no apparent correlations between flux density and redshift, while in X-rays the upper bound of count rate declines for the fainter sources. Therefore, at most wavelengths, flat-spectrum AGN are equally likely to be detected given the flux above the threshold of flux detection. In the X-rays, the extra flux dependence constraining the upper bound is due to the relative shallow depth of RASS. The bright X-ray sources at high redshift are either less likely to be detected by RASS, or the origin of X-ray emission can be different from other wavelengths.
}
\label{fig:s_to_z}
\end{figure}

\begin{figure*}
\begin{center}
    \includegraphics[width=1.00\textwidth]{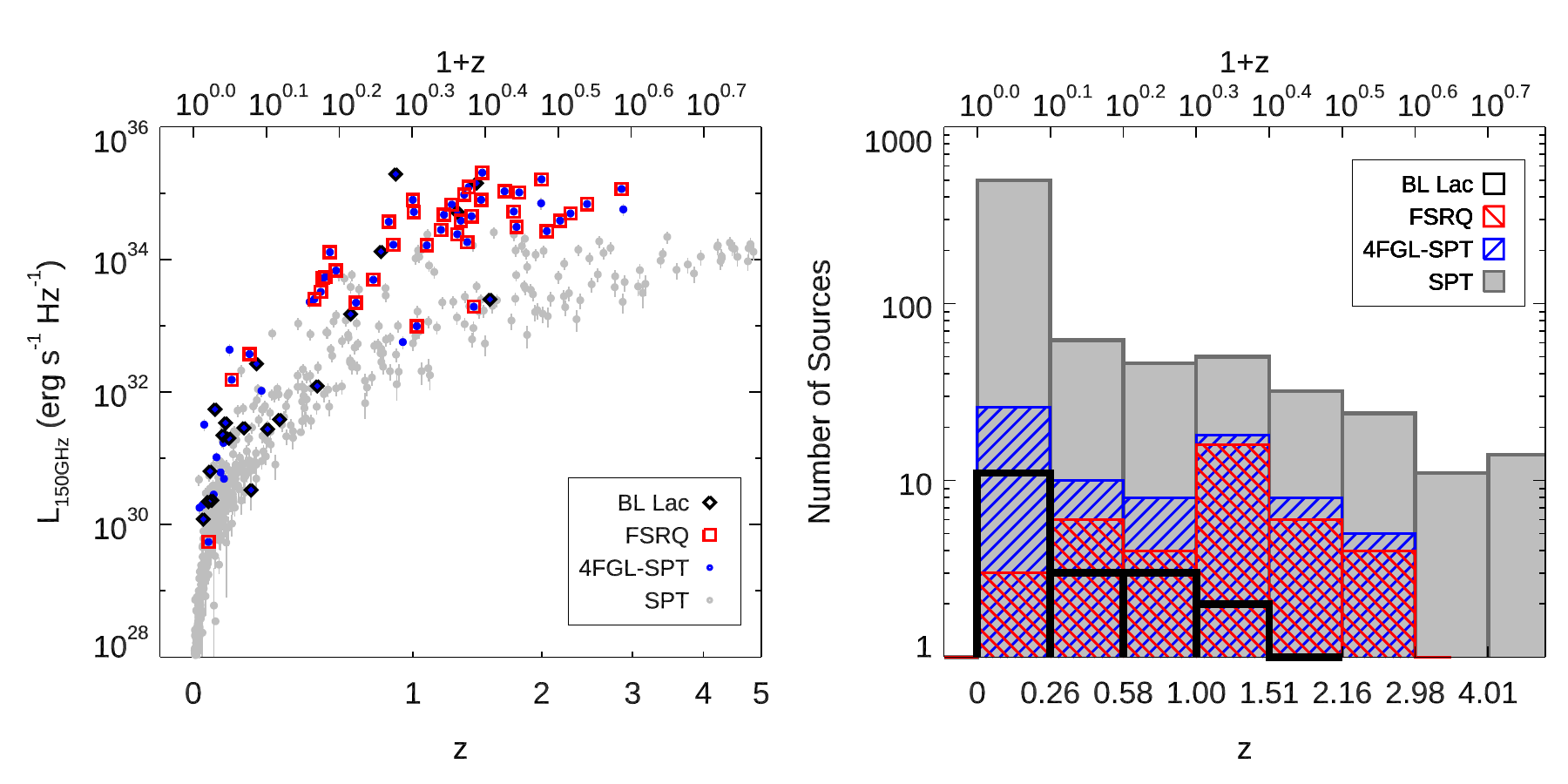}
\end{center}
\caption{
\textbf{Left Panel}: Specific luminosity in 150 GHz as the function of redshift. We have assumed a $\Lambda\mathrm{CDM}$ cosmology with $\Omega_M=0.27$, $\Omega_{\Lambda}=0.73$, and $H_0=71\ \mathrm{km\ s^{-1}\ Mpc^{-1}}$. \textit{Grey dots} represent all the SPT sources with redshift measurements. Those with 4FGL counterparts (4FGL-SPT) are covered by \textit{blue dots}. We also label the blazar type for each 4FGL-SPT source if it is classified in the catalog. Given the same redshift, mm-identified 4FGL sources (flat spectrum AGN) are brightest, where the specific luminosity is above the threshold of mm detection by orders of magnitude. This is resulting from the mixing of two types of sources (flat-spectrum and steep-spectrum AGN) and the selection bias of redshift data. Details are discussed in the discussion section (\ref{sec:results_discussion}). We can roughly distinguish the population of FSRQs and BL Lacs by their mm specific luminosity. 
\textbf{Right Panel}: Histogram of redshift distribution. The label colors are consistent with the left plot and redshifts are binned logarithmically. This histogram shows that the population of FSRQs and BL Lacs can be distinguished by their redshift measurements as well. FSRQs are mostly observed with higher redshift than BL Lacs. This distinction should come from the detection bias of optical spectrum measurement. In general, FSRQs have strong emission lines in optical broadband, while the broadband spectra of BL Lacs have weak emission lines or even featureless.
}
\label{fig:L_to_z}
\end{figure*}

\section{Results \& Discussion}
\label{sec:results_discussion}

Robust multi-wavelength counterpart identification to the sources in the 4FGL $\gamma$-ray catalog is crucial for understanding the $\gamma$-ray source population and the diffuse $\gamma$-ray background.
In this section we study and discuss the multi-wavelength properties of \Fermi $\gamma$-ray blazars, including the multi-wavelength flux correlation, multi-wavelength color analysis, redshift dependency, and implications for future surveys. 

We adopt two approaches to study the multi-wavelength properties of the 4FGL sources. 
The first approach directly adopts the multi-wavelength  associations derived from SPT+SUMSS and described in \autoref{sec:method}.
For the second approach, we divide the 4FGL sources into two groups -- previously associated 4FGL sources (ID-4FGL) and previously unassociated 4FGL sources with new mm identifications (UID-4FGL-SPT). For ID-4FGL sources (which have well-measured $\sim$arcsecond associated astrometry) we apply a simple angular-separation-based cross-matching with the external multi-wavelength catalogs. For UID-4FGL sources, i.e. the ones without a previous counterpart association, we use the counterpart associations found in this work using SPT+SUMSS. Both methods -- either using the provided cross-matching from the 4FGL catalog or our own independent cross-matching using SPT+SUMSS -- produce qualitatively similar results. For the rest of the figures in this paper, for the ID-4FGL sources we use the supplied position for multi-wavelength associations, and for the UID-4FGL sources we use the UID-4FGL-SPT position derived in this work.

The spectroscopic redshift data were acquired from the NED based on the refined source positions from the multi-wavelength associations. We found 75 sources (out of 239) with redshift measurements.

\subsection{The $\gamma$-ray Flux Correlation}
\label{sec:fluxcorrelation}

With the cross-matched multi-wavelength catalog in hand, we can study the flux correlation across bands for the 4FGL catalog. For each band, we calculate the Spearman's rank correlation coefficient ($r_{\mathrm{SC}}$) of the flux correlation and the results are shown in \autoref{fig:s_to_s}.
The UID-4FGL-SPT sources (black dots) are typically faint sources at all wavelengths studied here. The inclusion of these faint sources increases the completeness the 4FGL associations. 

Starting in the radio, we see that there is a slight correlation between the radio and $\gamma$-ray flux, with $r_{\mathrm{SC}}\sim0.3$.

The mm flux is well-correlated ($r_{\mathrm{SC}}\sim0.5$) to the $\gamma$-ray flux and can be parameterized by the following relation:
\begin{equation}
    \frac{S_{0.1-100\mathrm{GeV}}}{10^{-13}\ \mathrm{erg\ cm^{-2}\ s^{-1}}} = 0.8\left(\frac{S_{150\mathrm{GHz}}}{1\ \mathrm{mJy}}\right)^{1.1}
    \quad.
\end{equation}
Most of these sources are flat spectrum AGN, but a few sources have stronger radio emission (see 4FGL sources with $S_{\mathrm{150GHz}}>100\ \mathrm{mJy}$ outside the blue oval in \autoref{fig:s843_s150}). This may be partially related to source variability and may contribute to the radio flux correlating less well than the mm flux. 

The infrared flux is statistically as correlated as the mm ($r_{\mathrm{SC}}\sim0.5$). However, given the high density of the infrared \textit{WISE} catalog, we first needed to match the 4FGL to a SPT+SUMSS source, and then to \textit{WISE} in order to refine the positional uncertainty. Thus, this correlation may be biased towards the mm-detected sources.

The X-ray flux is largely uncorrelated to the $\gamma$-ray flux ($r_{\mathrm{SC}}\sim0$).  Naively, the lack of correlation between the $\gamma$-ray and X-ray is a surprise, given their proximity along the electromagnetic spectrum, particularly relative to the radio. 
This is likely due to the fact, that while gamma rays in the jet are always produce via inverse Compton scattering (of synchrotron photons in BL Lac, or external photons in FSRQs), the X-rays are generated via synchrotron in BL Lacs and via IC scattering in FSRQs.

Our study demonstrates that the mm band is the most efficient band to associate $\gamma$-ray blazars to multi-wavelength counterparts. As shown in the \textit{Method} \autoref{sec:method}, RASS's X-ray sensitivity is insufficient to detect all the 4FGL counterparts and the X-ray flux of 4FGL sources is observed to be uncorrelated to their $\gamma$-ray fluxes (\autoref{fig:s_to_s}). Thus, the RASS's X-ray catalog is highly incomplete in terms of associating 4FGL sources to multi-wavelength counterparts. The SUMSS radio associations have a high completeness, but often have multiple counterparts and are thus confused. \textit{WISE}, due to its high source density, produces a large number of spurious associations, and thus needs an accurate prior on the position to enable accurate source association. The mm catalog with arcminute resolution, such as the SPT catalog, provides 4FGL associations with both high completeness and purity (see \colorAutoref{tab:completeness} and \autoref{fig:dr_p_histogram}). 

\subsection{Previously Unidentified and Faint $\gamma$-ray Population}
\label{sec:faintpopulation}

As shown in \colorAutoref{fig:s_to_s}, $\gamma$-ray emission is correlated to the mm emission. Thus, the brighter 4FGL sources have a higher completeness of mm associations. The previously associated 4FGL sources (ID) typically have brighter multi-wavelength fluxes (see non-black sources in \colorAutoref{fig:s_to_s}), where 94\% of them are also SPT-identified. The previously unassociated 4FGL sources (UID) tend to be fainter ($S_{0.1-100\mathrm{GeV}}$$<$$10^{-11}$ $\mathrm{erg\ cm^{-2}\ s^{-1}}$) at longer wavelengths (see black sources in \colorAutoref{fig:s_to_s}), where only 56\% of them are SPT-identified. (See \autoref{tab:completeness} for a summary of these associations).
In addition, as can be seen in \autoref{fig:flux_alpha} (left), the remaining 4FGL sources without SPT-identification are mostly faint $\gamma$-ray sources ($S_{0.1-100\mathrm{GeV}} \lesssim 5 \times 10^{-12} \mathrm{erg\ cm^{-2}\ s^{-1}}$),  which might explain why they have remained unassociated thus far. 
Therefore, the identification of the remaining unassociated 4FGL blazars will be further completed by either deeper catalogs from upcoming surveys in mm wavelengths (e.g. SPT-3G, Simons Observatory, CMB-S4), or by dedicated pointed observations with e.g. ALMA, SMA, or NOEMA.

In \autoref{fig:s843_s150} we plot the radio versus  mm flux for all sources in the SPT catalog with a radio counterpart and highlight the sources with $\gamma$-ray counterparts. Some flat spectrum AGN with strong mm emission ($S_{\mathrm{150GHz}}>100\ \mathrm{mJy}$) are still undetected in $\gamma$-ray. 
Roughly half of them are $\gamma$-ray-quiet blazars and can also be found in CGRaBS \citep[see][]{2008ApJS..175...97H, 2017ApJ...851...33P} or ROMA-BZCAT \citep{2015Ap&SS.357...75M}. The rest of them are also blazar-like, but lack $\gamma$-ray detection. The $\gamma$-ray-quiet blazars are often associated with small Doppler factors and high disk dominance \citep{2017ApJ...851...33P}. 
As noted by several authors \citep{1995ApJ...446L..63D, 2017ApJ...851...33P, 2017MNRAS.469..255G}, there are two important selection effects which could make a luminous but high-redshift blazar not detected by \Fermi: (1) Luminous sources usually have the high-energy peaks of their SEDs shifted to lower energies, even as measured in the source frame, which is in addition to the effect of redshift itself. (2) 
These sources may be highly beamed at gamma rays and slightly misaligned making them currently undetected by Fermi-LAT.
Unfortunately we do not have the data necessary to resolve this issue at this time, but, for instance, a stacking analysis could be done in the future.

\subsection{Gamma-ray Blazars in Radio, mm, and Mid-IR}
\label{sec:fermi_wise}
By using multi-wavelength profiles and labeling the source type of associated 4FGL sources, we studied how the spectral characteristics influence the patterns in multiple parameter spaces.

First, we looked into the SPT-identified 4FGL sources within the SPT-SZ field in the radio, mm, and $\gamma$-ray band fluxes. FSRQs and BL Lacs have already been classified in the 4FGL catalog based on their optical spectra \citep{2020ApJS..247...33A}. As shown in the left panel of \autoref{fig:flux_alpha}, FSRQs are generally brighter than BL Lacs in mm wavelengths, but indistinguishable in $\gamma$-ray. The right panel demonstrates how the spectral indices vary with $\gamma$-ray flux, where the spectral index is defined as
\begin{equation}
    \alpha_{\mathrm{150GHz}}^{\mathrm{843MHz}} = \frac{\mathrm{log}\frac{S_{\mathrm{150GHz}}}{S_{\mathrm{843MHz}}}}{\mathrm{log}\frac{\mathrm{150GHz}}{\mathrm{843MHz}}}\mathrm{.}
\end{equation}
A flat spectrum from mm to radio corresponds to $\alpha_{\mathrm{150GHz}}^{\mathrm{843MHz}}=0$. As shown in the right panel of \autoref{fig:flux_alpha}, the radio spectral index of bright $\gamma$-ray sources tend to be more flat, presumably because the viewing angle is more aligned with the center of the jet thus the observed emission is dominated by the central engine of the jet. As $\gamma$-ray flux decreases below $\sim 5 \times 10^{-12} \mathrm{erg\ cm^{-2}\ s^{-1}}$, flat-spectrum and steep-spectrum AGN are heavily mixed together ($S_{843\mathrm{MHz}}$ and $S_{150\mathrm{GHz}}\lesssim100\ \mathrm{mJy}$ in \autoref{fig:s843_s150}), so the spectral index exhibits a higher scatter at lower $\gamma$-ray fluxes. 
Most of the sources ($<10^{-11}\ \mathrm{erg\ cm^{-2}\ s^{-1}}$) have a steeper (i.e. less flat) radio spectral index because they are less jet-dominated.
The radio emission from jets likely originates from an optically thick regime. 
Therefore, there could be additional radio emission if the surface is large. 
However, the jet is optically thin in the mm, where mm and $\gamma$-ray emission are likely to be produced co-spatially in the compact emission region in the jets \citep{2019ApJ...877...39M}. 
This is consistent with the higher flux correlation between mm and $\gamma$-ray than with the radio (see \autoref{fig:s_to_s}).

Next, we studied the 4FGL sources with joint mm and infrared counterparts in \textit{WISE} color space. The \textit{WISE} two-color diagram has clear patterns for the classes of \textit{WISE} objects (see also \citealt{2010AJ....140.1868W}). 
As shown in \autoref{fig:wise_color}, the background color contours represent the source number density of \textit{WISE} objects,  with density increasing from blue to red. Each population of \textit{WISE} objects is labeled on the diagram. For the SPT sources with \textit{WISE} counterparts, some of them have infrared colors indicative of being AGN-dominated, while others have a substantial component of starlight from an early-type host galaxy. Many other SPT sources which are not $\gamma$-ray emitters  have very blue [3.4]--[4.6] colors which is indicative of redshifts reaching up to around 1. The red dashed lines outline the region of \textit{WISE} Gamma-ray Strip (WGS, \citealt{2012ApJ...750..138M}) which indicates the location of the known blazars from ROMA-BZCAT in \textit{WISE} color space. The majority of SPT-identified blazars with $\gamma$-ray emission are located within both QSOs population and WGS. We find general consistency between our association results with \textit{WISE} source classification and WGS parameterization. Moreover, since the mm is efficient at associating $\gamma$-ray-loud flat spectrum AGN, it also provides a method to validate outliers from the WGS region and also pinpoint other possible populations that contain $\gamma$-ray emission along with infrared emission. 

\subsection{The Redshift Distribution of Gamma-ray Blazars}
With previous multi-wavelength associations, we obtained the redshifts for 31\% (75 out of 239 sources) of associated 4FGL sources by cross-matching NED database. An important caveat to the discussion below is that this sample is highly spectroscopically incomplete, and could be highly susceptible to selection biases. For instance, the sources with spectroscopic redshifts are presumably biased towards being optically bright and having strong emission lines. A future spectroscopic survey, or a dedicated targeted spectroscopic campaign, would be needed before any strong conclusions are drawn from this particular discussion.

In \autoref{fig:s_to_z} we investigate the redshift dependence on the multi-wavelength detection of 4FGL sources. With the exception of the X-ray band, the majority of multi-wavelength fluxes of 4FGL sources have no clear redshift dependence. In the left panel of \autoref{fig:L_to_z}, SPT-identified 4FGL sources (blue) are demonstrably more luminous in the mm-band than the sources not detected in 4FGL. 
In addition to the 4FGL-SPT sources, we also label the 4FGL classification of FSRQ (red) and BL Lac (black) and histogram the redshift distribution based on their types. In the right panel of \autoref{fig:L_to_z}, BL Lacs are mostly observed with lower redshift than FSRQs. This distinction may be due to an observational bias as BL Lacs generally have weak optical emission lines.

\section{Conclusion}
\label{sec:conclusion}
1. We have shown that the mm flux detected by SPT from flat-spectrum AGN 
is the most effective method currently available for identifying extragalactic 4FGL $\gamma$-ray sources and predicting the possible $\gamma$-ray emission. The SPT detection, combined with an accurate source position,
finds 4FGL sources with the highest completeness rate and, at the same time, the lowest
contamination rate. 

2. The effectiveness of the SPT+SUMSS selection has been demonstrated first by confirming the association of 94\% (199 out of 211 sources) of already-known 4FGL sources across the 2500 $\mathrm{deg^2}$ SPT-SZ survey field. It is then applied to identify 40 new sources for which 4FGL did not previously have a counterpart at lower energies (out of 71 previously unidentified sources within the SPT-SZ survey field). Deeper and wider mm surveys will soon be available which will greatly complete these 4FGL associations.

3. We have used multi-wavelength data to explain why SPT is more effective to find associations for $\gamma$-ray sources. Our interpretation is that the mm flux is closely correlated with the $\gamma$-ray flux \citep{2019ApJ...877...39M} because both have a common origin in the jet. 

4. SPT has shown its extraordinary ability to track jet-dominant AGN and find blazar-like objects. Therefore, SPT can also complete the sampling of $\gamma$-ray-quiet blazars. SPT has detected 60 bright mm emitters ($S_{\mathrm{150GHz}}>100\ \mathrm{mJy}$) which do not currently have any $\gamma$-ray detection. Roughly half of them are $\gamma$-ray-quiet blazars and can be found in CGRaBS \citep{2017ApJ...851...33P} or ROMA-BZCAT \citep{2015Ap&SS.357...75M}. The rest of them are also blazar-like sources and lack multi-wavelength detection. 

5. SPT-3G \citep{2014SPIE.9153E..1PB} will survey 1500 $\mathrm{deg^2}$ of southern sky $\times 10$ deeper and with polarization sensitivity. The DOE has recently begun planning a next-generation experiment (CMB-Stage 4, \citealt{2016arXiv161002743A}) to cover the entire extragalactic sky. Thus, this initial study enables us to prepare forecasts for the next-generation of SPT surveys and CMB experiments which will extend this technique to greater sensitivities and across the entire sky. With much more powerful data sets from SPT-3G and CMB-S4, we should be able to complete the association of the remaining unassociated 4FGL sources and also study the light curves of 4FGL blazars in both mm and $\gamma$-ray. The variability analysis will directly answer if mm and $\gamma$-ray emission from 4FGL blazars are intrinsically correlated, which is indicative of the radiation process inside relativistic jets. 

While deeper future surveys in the X-ray (e.g. eROSITA), the radio (e.g. VLASS, MeerKAT, ASKAP, SKA), and the mm regime (SPT-3G, CMB-S4) will help advance this work and improve statistics, we believe these general characteristics of the multi-wavelength associations to $\gamma$-ray catalogs will remain largely unchanged.


\section{Acknowledgements} 
The authors would like to thank Tom Crawford and Gil Holder for helpful conversations and crucial insights that greatly improved this paper.  This work was partially supported by NASA \textit{Fermi} Guest Observer Program number 101261. The SPT is supported by the NSF through grant OPP-1852617. J.D.V. acknowledges support from the NSF under grants AST-1715213 and AST-1716127. J.D.V. acknowledges support from an A. P. Sloan Foundation Fellowship. M.A.A. and J.D.V. acknowledge support from the Center for AstroPhysical Surveys at the National Center for Supercomputing Applications in Urbana, IL.
This research has made use of NASA's Astrophysics Data System Bibliographic Services. This research has made use of the NASA/IPAC Extragalactic Database (NED), which is operated by the Jet Propulsion Laboratory, California Institute of Technology, under contract with the National Aeronautics and Space Administration.

\bibliography{references}{}
\bibliographystyle{aasjournal}


\appendix
\begin{appendices}

\section{A. 4FGL-SPT Association Table}
\label{sec:long_table}

We list all 282 4FGL sources within 2500 $\mathrm{deg^2}$ SPT-SZ survey field. For each 4FGL source, all the SPT counterparts are listed on the right. The index is based on the sequence of the energy flux (100 MeV-100 GeV) from high to low. In this table, 4FGL sources with `*' represents that the source has no associated counterpart in the original 4FGL catalog. Most of SPT sources have detections greater than $4.5\sigma$ except those SPT sources whose name starts with `F' (for ``faint"), which represent the detection less than $4.5\sigma$ but greater than $1\sigma$. Recall that the uncertainty of mm flux includes mm variability, statistical and systematic noises, while the uncertainty of 4FGL energy flux is just the statistical noise. Source separations are listed in units of arcmin ($dr_1$) and $\sigma$ of the 4FGL positional uncertainty ($dr_2$).

\begin{center}
\setlength{\tabcolsep}{1.0pt}
\begin{longtable}{ l l r@{\,$\pm$\,}l l r@{\,$\pm$\,}l c c c}
\caption{282 4FGL Sources within 2500 $\mathrm{deg}^2$ SPT-SZ Survey Field} \label{tab:spt_4fgl} \\

\hline\hline
\multicolumn{1}{c}{\thead{\textbf{Index}\\\ }} & \multicolumn{1}{c}{\thead{\textbf{4FGL Name}\\\ }} & \multicolumn{2}{c}{\thead{$\boldsymbol{S_{0.1-100\mathrm{GeV}}}$\\$\boldsymbol{\mathrm{(10^{-13}\ erg\ cm^{-2}\ s^{-1})}}$}} & \multicolumn{1}{c}{\thead{\textbf{SPT Name}\\\ }} & \multicolumn{2}{c}{\thead{$\boldsymbol{S_{150\mathrm{GHz}}}$\\(\textbf{mJy})}} & \multicolumn{1}{c}{\thead{$\boldsymbol{dr_1}$\\(\textbf{arcmin})}} & \multicolumn{1}{c}{\thead{$\boldsymbol{dr_2}$\\($\boldsymbol{\sigma_{\mathrm{4FGL}}}$)}} & \multicolumn{1}{c}{\thead{$\boldsymbol{\log_{10}(p}$\textbf{-value)}\\\ }} \\
\hline
\endfirsthead

\multicolumn{3}{c}%
{{\bfseries \tablename\ \thetable{} -- continued from previous page}} \\
\hline\hline
\multicolumn{1}{c}{\thead{\textbf{Index}\\\ }} & \multicolumn{1}{c}{\thead{\textbf{4FGL Name}\\\ }} & \multicolumn{2}{c}{\thead{$\boldsymbol{S_{0.1-100\mathrm{GeV}}}$\\$\boldsymbol{\mathrm{(10^{-13}\ erg\ cm^{-2}\ s^{-1})}}$}} & \multicolumn{1}{c}{\thead{\textbf{SPT Name}\\\ }} & \multicolumn{2}{c}{\thead{$\boldsymbol{S_{150\mathrm{GHz}}}$\\(\textbf{mJy})}} & \multicolumn{1}{c}{\thead{$\boldsymbol{dr_1}$\\(\textbf{arcmin})}} & \multicolumn{1}{c}{\thead{$\boldsymbol{dr_2}$\\($\boldsymbol{\sigma_{\mathrm{4FGL}}}$)}} & \multicolumn{1}{c}{\thead{$\boldsymbol{\log_{10}(p}$\textbf{-value)}\\\ }} \\
\hline 
\endhead

\hline
\endfoot

\hline
\endlastfoot

0 & 4FGL J0538.8-4405  & 1993.0 & 26.0 & SPT-S J053850-4405.1 & 4843.0 & 970.2 & 0.09 & 0.36 & -8.54 \\
1 & 4FGL J2329.3-4955  & 1513.6 & 18.2 & SPT-S J232920-4955.6 & 522.1 & 104.6 & 0.41 & 1.44 & -5.95 \\
2 & 4FGL J0449.4-4350  & 1039.1 & 25.9 & SPT-S J044924-4350.0 & 119.3 & 23.9 & 0.29 & 1.12 & -5.51 \\
3 & 4FGL J2056.2-4714  & 675.1 & 12.5 & SPT-S J205616-4714.8 & 1424.9 & 285.4 & 0.63 & 1.55 & -6.38 \\
4 & 4FGL J0210.7-5101  & 474.0 & 10.2 & SPT-S J021045-5101.0 & 1504.4 & 301.4 & 0.32 & 0.69 & -7.17 \\
5 & 4FGL J0532.0-4827  & 448.8 & 15.1 & SPT-S J053158-4827.6 & 400.3 & 80.2 & 0.29 & 0.77 & -6.10 \\
6 & 4FGL J2139.4-4235  & 424.3 & 14.5 & SPT-S J213924-4235.4 & 154.0 & 30.9 & 0.11 & 0.30 & -6.44 \\
7 & 4FGL J0245.9-4650  & 374.9 & 9.2 & SPT-S J024600-4651.2 & 1036.5 & 207.6 & 0.47 & 0.90 & -6.15 \\
8 & 4FGL J0334.2-4008  & 349.3 & 10.7 & SPT-S J033413-4008.4 & 1068.5 & 214.1 & 0.27 & 0.63 & -6.76 \\
9 & 4FGL J2009.4-4849  & 335.0 & 13.1 & SPT-S J200925-4849.7 & 457.8 & 91.7 & 0.31 & 0.90 & -6.15 \\
10 & 4FGL J2141.7-6410  & 313.4 & 10.0 & SPT-S J214145-6411.1 & 157.3 & 31.5 & 0.50 & 0.97 & -5.19 \\
11 & 4FGL J0309.9-6058  & 277.7 & 8.7 & SPT-S J030956-6058.6 & 563.7 & 112.9 & 0.32 & 0.54 & -6.31 \\
12 & 4FGL J2241.7-5236  & 266.1 & 12.8 & - & \multicolumn{1}{r}{-} &  & - & - & - \\
13 & 4FGL J0228.3-5547  & 262.4 & 16.2 & SPT-S J022821-5546.0 & 138.2 & 27.7 & 1.53 & 2.79 & -4.16 \\
14 & 4FGL J0516.7-6207  & 232.7 & 10.4 & SPT-S J051644-6207.0 & 510.1 & 102.2 & 0.49 & 1.08 & -5.74 \\
15 & 4FGL J2328.3-4036  & 226.9 & 11.0 & SPT-S J232818-4035.1 & 387.7 & 77.7 & 1.06 & 1.58 & -4.97 \\
16 & 4FGL J0526.2-4830  & 210.4 & 11.3 & SPT-S J052616-4830.6 & 236.1 & 47.3 & 0.23 & 0.39 & -6.01 \\
17 & 4FGL J0209.3-5228  & 187.6 & 9.7 & SPT-S J020921-5229.3 & 24.0 & 4.9 & 0.66 & 1.40 & -4.03 \\
18 & 4FGL J0543.9-5531  & 175.1 & 11.0 & SPT-S J054357-5532.1 & 17.3 & 3.6 & 0.34 & 0.83 & -4.44 \\
19 & 4FGL J0437.2-4715  & 167.7 & 6.0 & - & \multicolumn{1}{r}{-} &  & - & - & - \\
20 & 4FGL J0051.2-6242  & 162.7 & 10.8 & SPT-S J005115-6241.9 & 11.3 & 2.5 & 0.34 & 0.79 & -4.27 \\
21 & 4FGL J0030.3-4224  & 161.4 & 6.4 & SPT-S J003017-4224.8 & 347.6 & 69.6 & 0.98 & 1.09 & -4.89 \\
22 & 4FGL J0455.7-4617  & 160.2 & 7.3 & SPT-S J045550-4615.9 & 1043.0 & 209.0 & 1.45 & 1.69 & -5.24 \\
23 & 4FGL J2126.3-4605  & 149.3 & 6.6 & SPT-S J212630-4605.6 & 560.1 & 112.2 & 1.40 & 1.34 & -4.96 \\
24 & 4FGL J2039.5-5617* & 149.3 & 8.3 & - & \multicolumn{1}{r}{-} &  & - & - & - \\
25 & 4FGL J2103.8-6233  & 145.3 & 8.5 & SPT-S J210337-6232.4 & 81.6 & 16.4 & 1.61 & 3.02 & -3.78 \\
26 & 4FGL J0143.7-5846  & 143.1 & 8.6 & SPT-S J014347-5845.7 & 11.7 & 2.7 & 0.59 & 1.16 & -3.81 \\
27 & 4FGL J2052.2-5533  & 132.6 & 6.5 & SPT-S J205213-5533.1 & 36.6 & 7.4 & 0.68 & 0.77 & -4.22 \\
28 & 4FGL J2135.3-5006  & 132.5 & 6.2 & SPT-S J213521-5006.7 & 104.6 & 21.0 & 0.71 & 0.77 & -4.65 \\
29 & 4FGL J0101.1-6422  & 129.9 & 5.4 & - & \multicolumn{1}{r}{-} &  & - & - & - \\
30 & 4FGL J0236.8-6136  & 125.9 & 6.7 & SPT-S J023652-6136.1 & 313.8 & 62.9 & 0.70 & 0.91 & -5.11 \\
31 & 4FGL J0433.6-6030  & 122.0 & 6.5 & SPT-S J043333-6030.0 & 266.9 & 53.5 & 0.73 & 0.89 & -5.03 \\
32 & 4FGL J2324.7-4041  & 116.4 & 10.2 & SPT-S J232444-4040.8 & 35.0 & 7.2 & 0.22 & 0.44 & -5.19 \\
33 & 4FGL J2325.4-4800  & 116.2 & 6.8 & SPT-S J232527-4800.2 & 188.8 & 37.8 & 0.57 & 0.94 & -5.08 \\
34 & 4FGL J0303.6-6211  & 112.5 & 6.0 & SPT-S J030350-6211.4 & 847.4 & 169.8 & 1.04 & 1.20 & -5.38 \\
& & \multicolumn{1}{r}{} & & SPT-S J030337-6214.6 & 22.5 & 4.6 & 3.32 & 3.75 & -2.59 \\
35 & 4FGL J2235.3-4836  & 105.3 & 5.4 & SPT-S J223513-4835.9 & 534.3 & 107.0 & 1.45 & 1.37 & -4.89 \\
36 & 4FGL J0507.7-6104  & 102.4 & 7.3 & SPT-S J050754-6104.6 & 249.5 & 50.0 & 1.29 & 1.26 & -4.52 \\
37 & 4FGL J0133.1-5201  & 99.2 & 5.7 & SPT-S J013305-5200.1 & 204.5 & 41.0 & 1.39 & 1.28 & -4.39 \\
38 & 4FGL J0343.2-6444  & 98.7 & 6.0 & SPT-S J034320-6442.8 & 20.9 & 4.3 & 1.36 & 1.88 & -3.32 \\
39 & 4FGL J0602.8-4019  & 98.5 & 8.7 & SPT-S J060250-4018.7 & 26.3 & 5.4 & 1.02 & 1.68 & -3.69 \\
40 & 4FGL J0424.9-5331  & 97.2 & 5.2 & SPT-S J042504-5331.8 & 128.1 & 25.7 & 0.70 & 0.76 & -4.79 \\
41 & 4FGL J2007.9-4432  & 92.3 & 7.6 & SPT-S J200755-4434.6 & 61.3 & 12.3 & 2.51 & 1.85 & -3.31 \\
42 & 4FGL J0515.6-4556  & 92.2 & 7.3 & SPT-S J051544-4556.6 & 354.6 & 71.0 & 1.17 & 1.03 & -4.78 \\
43 & 4FGL J2221.5-5225  & 88.8 & 8.2 & SPT-S J222129-5225.5 & 16.8 & 3.6 & 0.68 & 1.05 & -3.84 \\
44 & 4FGL J0438.9-4521  & 85.5 & 6.5 & SPT-S J043900-4522.3 & 278.3 & 55.8 & 0.87 & 0.83 & -4.93 \\
45 & 4FGL J2336.6-4115  & 84.3 & 5.6 & SPT-S J233633-4115.3 & 362.1 & 72.6 & 1.06 & 0.90 & -4.96 \\
46 & 4FGL J2022.3-4513  & 83.5 & 6.1 & SPT-S J202226-4513.4 & 184.9 & 37.1 & 0.90 & 1.00 & -4.76 \\
47 & 4FGL J2207.5-5346  & 83.2 & 5.1 & SPT-S J220743-5346.5 & 559.6 & 112.1 & 1.40 & 1.02 & -5.05 \\
48 & 4FGL J2347.9-5106* & 82.3 & 5.3 & - & \multicolumn{1}{r}{-} &  & - & - & - \\
49 & 4FGL J0647.7-6058  & 81.0 & 6.0 & SPT-S J064740-6058.0 & 86.6 & 17.4 & 0.74 & 0.65 & -4.47 \\
50 & 4FGL J0331.3-6156  & 79.7 & 5.8 & SPT-S J033118-6155.3 & 21.2 & 4.4 & 0.97 & 1.41 & -3.63 \\
51 & 4FGL J2321.7-6438  & 77.5 & 5.2 & SPT-S J232142-6438.1 & 39.2 & 8.0 & 0.56 & 0.68 & -4.47 \\
52 & 4FGL J2357.8-5311  & 75.9 & 6.1 & SPT-S J235753-5311.2 & 971.8 & 194.7 & 0.41 & 0.29 & -6.27 \\
53 & 4FGL J0608.9-5456  & 69.6 & 5.5 & SPT-S J060848-5456.6 & 175.2 & 35.1 & 1.21 & 0.64 & -4.46 \\
54 & 4FGL J0021.9-5140  & 69.0 & 5.3 & SPT-S J002159-5140.4 & 17.5 & 3.7 & 0.06 & 0.08 & -5.94 \\
55 & 4FGL J0540.8-5415  & 68.8 & 6.6 & SPT-S J054045-5418.3 & 346.6 & 69.4 & 3.00 & 1.82 & -3.94 \\
56 & 4FGL J0156.9-5301  & 68.0 & 6.7 & SPT-S J015657-5302.0 & 9.8 & 2.3 & 0.35 & 0.47 & -4.22 \\
57 & 4FGL J0506.9-5435  & 67.9 & 7.8 & SPT-S J050658-5435.0 & 6.4 & 1.6 & 0.20 & 0.40 & -4.45 \\
58 & 4FGL J0244.6-5819  & 67.0 & 6.4 & SPT-S J024439-5819.8 & 12.9 & 2.8 & 0.41 & 0.62 & -4.10 \\
59 & 4FGL J0628.8-6250  & 67.0 & 5.3 & SPT-S J062857-6248.8 & 75.8 & 15.2 & 1.70 & 1.69 & -3.77 \\
60 & 4FGL J0004.4-4737  & 65.1 & 4.7 & SPT-S J000435-4736.3 & 433.9 & 86.9 & 1.89 & 1.47 & -4.53 \\
61 & 4FGL J0314.3-5103  & 63.9 & 5.2 & SPT-S J031425-5104.4 & 86.4 & 17.4 & 1.27 & 1.42 & -4.04 \\
62 & 4FGL J2251.5-4928  & 63.1 & 6.8 & SPT-S J225129-4929.1 & 11.2 & 2.5 & 1.03 & 1.61 & -3.26 \\
63 & 4FGL J0450.3-4419  & 62.9 & 10.0 & SPT-S J045001-4418.2 & 86.9 & 17.5 & 3.19 & 1.24 & -3.24 \\
64 & 4FGL J2322.8-4916  & 62.0 & 7.5 & SPT-S J232254-4916.6 & 8.4 & 2.1 & 0.76 & 1.15 & -3.42 \\
65 & 4FGL J0316.2-6437  & 61.6 & 6.0 & SPT-S J031613-6437.5 & 9.2 & 2.1 & 0.19 & 0.25 & -4.75 \\
66 & 4FGL J0231.2-4745  & 61.5 & 5.0 & SPT-S J023112-4746.1 & 182.2 & 36.5 & 0.75 & 0.29 & -4.85 \\
67 & 4FGL J0451.8-4651  & 61.1 & 5.8 & SPT-S J045153-4653.2 & 74.0 & 14.9 & 1.96 & 1.34 & -3.60 \\
68 & 4FGL J0413.1-5332  & 60.3 & 5.5 & SPT-S J041313-5331.9 & 17.6 & 3.7 & 1.42 & 0.95 & -3.27 \\
& & \multicolumn{1}{r}{} & & SPT-S J041303-5333.8 & 7.8 & 1.9 & 1.06 & 0.70 & -3.15 \\
69 & 4FGL J2132.0-5418  & 58.9 & 5.2 & SPT-S J213208-5420.4 & 98.4 & 19.7 & 1.75 & 0.94 & -3.77 \\
70 & 4FGL J0032.3-5522  & 58.4 & 4.5 & SPT-S J003210-5522.5 & 25.4 & 5.3 & 1.20 & 1.07 & -3.51 \\
71 & 4FGL J2240.3-5241* & 58.1 & 12.2 & SPT-S J224016-5241.3 & 18.0 & 3.8 & 0.59 & 0.79 & -4.00 \\
72 & 4FGL J0157.7-4614  & 57.6 & 7.3 & SPT-S J015751-4614.4 & 119.6 & 24.0 & 1.24 & 0.80 & -4.25 \\
73 & 4FGL J2317.4-4533  & 57.2 & 5.1 & SPT-S J231731-4533.9 & 11.4 & 2.6 & 1.01 & 1.07 & -3.30 \\
74 & 4FGL J0525.6-6013  & 57.0 & 5.7 & FSPT-S J052542-6013.6 & 4.3 & 0.9 & 0.14 & 0.19 & -4.61 \\
75 & 4FGL J0050.0-5736  & 55.3 & 5.0 & SPT-S J004959-5738.4 & 449.6 & 90.1 & 1.58 & 1.22 & -4.64 \\
76 & 4FGL J0556.2-4352  & 52.9 & 5.3 & SPT-S J055617-4351.9 & 40.5 & 8.2 & 0.23 & 0.22 & -5.17 \\
77 & 4FGL J0059.4-5654* & 52.3 & 6.9 & FSPT-S J005926-5657.1 & 14.5 & 2.9 & 2.61 & 1.58 & -2.53 \\
78 & 4FGL J0608.1-6028  & 51.5 & 7.7 & SPT-S J060755-6031.8 & 193.3 & 38.7 & 3.58 & 1.55 & -3.57 \\
79 & 4FGL J0335.1-4459  & 51.4 & 5.0 & SPT-S J033514-4459.5 & 17.6 & 3.7 & 1.01 & 1.01 & -3.51 \\
80 & 4FGL J0514.6-4408  & 50.9 & 3.5 & SPT-S J051422-4403.0 & 16.5 & 3.6 & 5.73 & 3.25 & -1.96 \\
81 & 4FGL J2250.4-4206  & 50.9 & 4.7 & SPT-S J225022-4206.2 & 55.9 & 11.3 & 0.73 & 0.81 & -4.35 \\
82 & 4FGL J0325.5-5635  & 50.8 & 5.0 & SPT-S J032522-5635.6 & 27.3 & 5.6 & 1.17 & 1.42 & -3.64 \\
83 & 4FGL J2315.6-5018  & 50.3 & 5.1 & SPT-S J231545-5018.6 & 424.2 & 85.0 & 0.96 & 0.80 & -5.07 \\
84 & 4FGL J0310.6-5017  & 49.6 & 5.8 & FSPT-S J031034-5016.5 & 4.8 & 1.0 & 0.95 & 1.22 & -2.94 \\
85 & 4FGL J2133.1-6432* & 48.7 & 4.8 & - & \multicolumn{1}{r}{-} &  & - & - & - \\
86 & 4FGL J0357.0-4955  & 47.3 & 4.3 & SPT-S J035658-4955.7 & 84.9 & 17.0 & 0.62 & 0.58 & -4.68 \\
87 & 4FGL J0519.6-4544  & 46.7 & 7.8 & SPT-S J051948-4546.6 & 1163.6 & 233.1 & 2.83 & 1.19 & -5.00 \\
& & \multicolumn{1}{r}{} & & SPT-S J051927-4545.8 & 356.7 & 71.5 & 2.10 & 0.97 & -4.25 \\
& & \multicolumn{1}{r}{} & & SPT-S J052005-4547.3 & 250.4 & 50.2 & 5.65 & 2.32 & -3.32 \\
88 & 4FGL J0433.7-5725  & 46.1 & 4.6 & SPT-S J043343-5726.4 & 5.9 & 1.6 & 0.80 & 0.85 & -3.20 \\
89 & 4FGL J2034.8-4200  & 44.6 & 5.4 & FSPT-S J203451-4200.4 & 8.1 & 1.6 & 0.43 & 0.54 & -3.85 \\
90 & 4FGL J2231.0-4416  & 43.2 & 4.2 & SPT-S J223056-4416.4 & 359.9 & 72.1 & 1.11 & 0.54 & -4.85 \\
91 & 4FGL J2040.0-5737  & 42.9 & 5.7 & SPT-S J204000-5735.2 & 111.1 & 22.3 & 1.93 & 1.26 & -3.82 \\
92 & 4FGL J0537.7-5717  & 42.9 & 4.7 & SPT-S J053749-5718.3 & 15.9 & 3.3 & 1.37 & 1.12 & -3.18 \\
93 & 4FGL J0253.2-5441  & 42.3 & 4.1 & SPT-S J025329-5441.8 & 1121.4 & 224.6 & 2.37 & 1.31 & -4.90 \\
& & \multicolumn{1}{r}{} & & SPT-S J025307-5441.1 & 2.2 & 0.6 & 0.98 & 0.52 & -2.84 \\
94 & 4FGL J0025.7-4801  & 41.9 & 4.6 & SPT-S J002545-4803.8 & 129.8 & 26.0 & 2.70 & 1.81 & -3.60 \\
95 & 4FGL J2329.3-4733  & 41.6 & 5.7 & SPT-S J232918-4730.3 & 712.6 & 142.8 & 3.10 & 1.71 & -4.44 \\
96 & 4FGL J0625.8-5441  & 41.3 & 5.5 & SPT-S J062552-5438.9 & 84.5 & 17.0 & 2.64 & 0.80 & -3.43 \\
97 & 4FGL J0022.0-5921* & 41.1 & 4.2 & FSPT-S J002127-5919.8 & 3.9 & 0.8 & 4.95 & 1.72 & -1.53 \\
98 & 4FGL J2342.4-4739* & 40.3 & 4.6 & - & \multicolumn{1}{r}{-} &  & - & - & - \\
99 & 4FGL J2056.4-5922* & 39.5 & 4.9 & SPT-S J205624-5917.4 & 11.1 & 2.4 & 4.66 & 2.12 & -2.01 \\
100 & 4FGL J2333.1-5527* & 39.4 & 4.4 & - & \multicolumn{1}{r}{-} &  & - & - & - \\
101 & 4FGL J0509.9-6417  & 38.2 & 4.5 & SPT-S J050957-6417.9 & 7.2 & 1.9 & 0.57 & 0.53 & -3.63 \\
102 & 4FGL J2358.0-4601  & 38.0 & 4.4 & SPT-S J235802-4555.2 & 319.3 & 64.0 & 5.85 & 3.58 & -3.33 \\
103 & 4FGL J0303.4-5232  & 37.5 & 4.2 & SPT-S J030328-5234.5 & 35.1 & 7.1 & 2.51 & 0.71 & -3.04 \\
104 & 4FGL J0643.2-5356  & 36.5 & 4.9 & SPT-S J064319-5358.7 & 77.6 & 15.6 & 2.61 & 1.54 & -3.36 \\
105 & 4FGL J2130.4-4241  & 36.5 & 5.2 & FSPT-S J213017-4244.4 & 1.7 & 0.3 & 3.33 & 1.24 & -1.73 \\
& & \multicolumn{1}{r}{} & & FSPT-S J213017-4243.3 & 2.5 & 0.5 & 2.29 & 0.88 & -2.05 \\
106 & 4FGL J0051.5-4220  & 36.4 & 4.8 & SPT-S J005109-4226.5 & 247.6 & 49.6 & 6.99 & 1.89 & -3.03 \\
107 & 4FGL J0132.8-4413  & 36.2 & 4.0 & SPT-S J013306-4414.4 & 19.9 & 4.2 & 3.11 & 2.31 & -2.62 \\
108 & 4FGL J0606.5-4730  & 36.0 & 4.9 & SPT-S J060635-4729.7 & 99.7 & 20.0 & 0.54 & 0.39 & -4.89 \\
109 & 4FGL J0003.1-5248  & 36.0 & 4.9 & - & \multicolumn{1}{r}{-} &  & - & - & - \\
110 & 4FGL J0226.5-4441  & 35.9 & 5.0 & FSPT-S J022627-4441.1 & 4.5 & 0.9 & 1.44 & 1.13 & -2.63 \\
& & \multicolumn{1}{r}{} & & FSPT-S J022638-4441.3 & 7.4 & 1.5 & 0.61 & 0.47 & -3.61 \\
111 & 4FGL J2258.4-5524  & 35.9 & 4.0 & SPT-S J225819-5525.5 & 14.3 & 3.0 & 1.16 & 0.81 & -3.30 \\
112 & 4FGL J0610.9-6054  & 35.1 & 6.6 & SPT-S J061030-6058.6 & 140.6 & 28.2 & 4.69 & 2.28 & -3.25 \\
113 & 4FGL J0049.4-5402  & 35.0 & 4.0 & SPT-S J004948-5402.7 & 16.4 & 3.5 & 3.30 & 2.00 & -2.48 \\
114 & 4FGL J2054.1-4054* & 35.0 & 5.2 & FSPT-S J205409-4050.4 & 1.3 & 0.3 & 3.83 & 0.82 & -1.59 \\
& & \multicolumn{1}{r}{} & & FSPT-S J205422-4051.3 & 5.0 & 1.0 & 3.94 & 0.80 & -1.72 \\
115 & 4FGL J0604.1-4816  & 34.8 & 5.2 & SPT-S J060409-4817.4 & 8.3 & 2.0 & 0.73 & 0.78 & -3.46 \\
116 & 4FGL J0557.5-4452* & 34.3 & 5.7 & - & \multicolumn{1}{r}{-} &  & - & - & - \\
117 & 4FGL J0528.7-5920  & 34.3 & 4.2 & SPT-S J052846-5919.8 & 11.7 & 2.5 & 0.80 & 0.64 & -3.54 \\
118 & 4FGL J0146.9-5202  & 34.2 & 4.1 & SPT-S J014648-5202.5 & 93.3 & 18.7 & 1.12 & 0.80 & -4.22 \\
119 & 4FGL J0001.6-4156  & 33.9 & 4.9 & FSPT-S J000133-4155.4 & 2.1 & 0.4 & 1.72 & 0.83 & -2.46 \\
120 & 4FGL J0535.1-5422* & 33.8 & 5.7 & - & \multicolumn{1}{r}{-} &  & - & - & - \\
121 & 4FGL J0138.5-4613  & 33.5 & 4.1 & SPT-S J013834-4614.2 & 30.9 & 6.3 & 1.22 & 0.94 & -3.50 \\
122 & 4FGL J0525.4-4600  & 33.4 & 5.1 & SPT-S J052532-4559.8 & 5.7 & 1.6 & 0.82 & 0.39 & -3.15 \\
& & \multicolumn{1}{r}{} & & SPT-S J052531-4557.9 & 17.7 & 3.8 & 2.46 & 1.22 & -2.70 \\
123 & 4FGL J0034.0-4116  & 33.1 & 4.7 & SPT-S J003404-4116.4 & 77.8 & 15.6 & 0.29 & 0.14 & -5.32 \\
124 & 4FGL J2043.9-4802* & 32.2 & 3.4 & - & \multicolumn{1}{r}{-} &  & - & - & - \\
125 & 4FGL J2159.8-4751  & 32.1 & 4.7 & SPT-S J215958-4751.9 & 43.3 & 8.8 & 1.59 & 0.41 & -3.67 \\
& & \multicolumn{1}{r}{} & & SPT-S J215859-4748.9 & 8.8 & 2.2 & 8.85 & 2.29 & -1.44 \\
126 & 4FGL J2343.7-5624  & 32.0 & 4.3 & SPT-S J234327-5626.2 & 104.8 & 21.0 & 3.09 & 1.40 & -3.39 \\
127 & 4FGL J0617.6-4028* & 31.6 & 5.8 & SPT-S J061646-4021.7 & 32.8 & 6.7 & 12.10 & 3.08 & -1.61 \\
128 & 4FGL J0200.3-4109  & 31.5 & 4.7 & - & \multicolumn{1}{r}{-} &  & - & - & - \\
129 & 4FGL J0116.2-6153  & 31.4 & 4.2 & SPT-S J011619-6153.7 & 16.0 & 3.4 & 0.56 & 0.70 & -3.99 \\
130 & 4FGL J2056.4-4904  & 30.8 & 6.4 & SPT-S J205614-4904.1 & 12.5 & 2.8 & 2.52 & 0.45 & -2.59 \\
& & \multicolumn{1}{r}{} & & SPT-S J205714-4901.3 & 29.6 & 6.1 & 7.91 & 1.51 & -1.93 \\
131 & 4FGL J0533.1-6119  & 30.7 & 5.3 & SPT-S J053435-6106.2 & 237.6 & 47.6 & 16.84 & 3.01 & -2.21 \\
& & \multicolumn{1}{r}{} & & SPT-S J053304-6115.9 & 6.1 & 1.7 & 3.41 & 0.58 & -1.93 \\
& & \multicolumn{1}{r}{} & & SPT-S J053047-6115.4 & 47.0 & 9.5 & 17.25 & 2.66 & -1.57 \\
132 & 4FGL J0647.7-4418  & 30.6 & 4.8 & FSPT-S J064744-4419.7 & 4.4 & 0.9 & 1.50 & 0.91 & -2.39 \\
133 & 4FGL J2143.0-5501  & 30.5 & 5.6 & SPT-S J214121-5504.4 & 39.3 & 8.0 & 14.66 & 0.67 & -1.74 \\
134 & 4FGL J2319.1-4207  & 30.2 & 4.2 & SPT-S J231905-4206.7 & 92.4 & 18.5 & 0.33 & 0.36 & -5.30 \\
135 & 4FGL J2237.6-5126* & 29.9 & 5.2 & SPT-S J223825-5114.3 & 61.1 & 12.3 & 13.71 & 2.77 & -1.87 \\
& & \multicolumn{1}{r}{} & & SPT-S J223553-5131.8 & 21.3 & 4.5 & 17.52 & 3.45 & -1.20 \\
136 & 4FGL J2117.1-5307* & 29.7 & 4.4 & SPT-S J211704-5306.8 & 3.3 & 1.3 & 0.80 & 0.63 & -3.06 \\
& & \multicolumn{1}{r}{} & & SPT-S J211701-5306.7 & 7.0 & 1.8 & 1.12 & 0.90 & -2.99 \\
137 & 4FGL J0231.2-5754  & 29.6 & 4.2 & SPT-S J023108-5755.1 & 52.6 & 10.6 & 0.98 & 0.69 & -4.17 \\
138 & 4FGL J2100.0-4356* & 29.6 & 4.8 & - & \multicolumn{1}{r}{-} &  & - & - & - \\
139 & 4FGL J2017.5-4113  & 29.6 & 4.8 & SPT-S J201729-4115.3 & 16.4 & 3.5 & 1.56 & 1.24 & -3.08 \\
140 & 4FGL J0622.4-6433  & 29.4 & 5.4 & SPT-S J062307-6436.4 & 357.7 & 71.7 & 5.25 & 1.15 & -3.38 \\
& & \multicolumn{1}{r}{} & & SPT-S J062336-6434.6 & 14.3 & 3.1 & 7.46 & 1.33 & -1.74 \\
& & \multicolumn{1}{r}{} & & SPT-S J062020-6438.8 & 13.7 & 3.0 & 14.63 & 2.17 & -1.30 \\
141 & 4FGL J2001.9-5737  & 29.0 & 4.4 & SPT-S J200204-5736.6 & 6.2 & 1.6 & 1.47 & 1.24 & -2.65 \\
142 & 4FGL J0014.1-5022  & 28.8 & 4.2 & FSPT-S J001411-5022.5 & 1.8 & 0.4 & 0.33 & 0.18 & -3.83 \\
143 & 4FGL J2159.6-4620* & 28.8 & 5.0 & - & \multicolumn{1}{r}{-} &  & - & - & - \\
144 & 4FGL J0156.8-4744  & 28.7 & 4.1 & SPT-S J015645-4744.2 & 7.5 & 1.9 & 1.01 & 0.68 & -3.11 \\
145 & 4FGL J2313.9-4501  & 28.7 & 5.9 & SPT-S J231408-4455.8 & 78.0 & 15.7 & 5.71 & 1.35 & -2.67 \\
146 & 4FGL J0251.5-5958  & 28.7 & 4.0 & SPT-S J025125-6000.1 & 156.6 & 31.4 & 2.10 & 1.03 & -4.04 \\
& & \multicolumn{1}{r}{} & & SPT-S J025202-5953.6 & 4.8 & 1.4 & 5.84 & 2.86 & -1.52 \\
147 & 4FGL J2326.9-4130* & 28.3 & 7.2 & SPT-S J232625-4140.2 & 43.9 & 8.9 & 11.35 & 3.60 & -1.80 \\
& & \multicolumn{1}{r}{} & & FSPT-S J232719-4134.5 & 3.1 & 0.6 & 5.80 & 1.69 & -1.35 \\
148 & 4FGL J0626.4-4259  & 28.2 & 4.8 & FSPT-S J062636-4258.1 & 3.7 & 0.8 & 2.07 & 1.49 & -2.29 \\
149 & 4FGL J0657.4-4658* & 27.8 & 4.7 & - & \multicolumn{1}{r}{-} &  & - & - & - \\
150 & 4FGL J0445.1-6012  & 27.6 & 3.9 & SPT-S J044500-6014.9 & 16.5 & 3.5 & 2.43 & 1.93 & -2.66 \\
151 & 4FGL J2234.2-4156* & 26.8 & 3.9 & SPT-S J223415-4156.9 & 9.7 & 2.3 & 0.49 & 0.38 & -3.87 \\
152 & 4FGL J2024.8-6459  & 26.7 & 4.5 & SPT-S J202445-6458.6 & 95.2 & 19.1 & 0.83 & 0.54 & -4.48 \\
153 & 4FGL J2115.6-4938  & 26.4 & 4.2 & SPT-S J211545-4938.8 & 6.9 & 1.8 & 1.46 & 1.05 & -2.76 \\
154 & 4FGL J0646.4-5455  & 26.3 & 3.2 & SPT-S J064628-5451.2 & 27.7 & 5.7 & 4.17 & 2.82 & -2.49 \\
155 & 4FGL J0503.1-6045  & 26.2 & 6.6 & SPT-S J050401-6049.8 & 59.6 & 12.0 & 7.70 & 1.82 & -2.34 \\
& & \multicolumn{1}{r}{} & & SPT-S J050335-6058.4 & 10.0 & 2.4 & 13.67 & 3.06 & -1.07 \\
156 & 4FGL J2041.1-6138* & 26.1 & 4.3 & SPT-S J204111-6139.8 & 16.2 & 3.4 & 1.73 & 0.87 & -3.03 \\
157 & 4FGL J0328.4-4736* & 25.4 & 3.7 & SPT-S J032842-4739.6 & 9.1 & 2.2 & 4.12 & 2.20 & -2.00 \\
158 & 4FGL J0651.9-4330* & 25.0 & 5.5 & - & \multicolumn{1}{r}{-} &  & - & - & - \\
159 & 4FGL J0048.6-6347* & 25.0 & 3.7 & - & \multicolumn{1}{r}{-} &  & - & - & - \\
160 & 4FGL J2046.8-4258  & 25.0 & 4.5 & SPT-S J204644-4257.2 & 7.7 & 2.0 & 1.82 & 1.02 & -2.51 \\
& & \multicolumn{1}{r}{} & & SPT-S J204643-4256.7 & 3.6 & 1.5 & 2.13 & 1.22 & -2.09 \\
161 & 4FGL J0442.0-5432* & 24.9 & 3.9 & SPT-S J044230-5431.7 & 52.3 & 10.5 & 4.15 & 1.87 & -2.81 \\
162 & 4FGL J0416.2-4353  & 24.9 & 4.1 & SPT-S J041613-4350.9 & 9.8 & 2.3 & 2.44 & 0.74 & -2.42 \\
& & \multicolumn{1}{r}{} & & SPT-S J041642-4401.9 & 7.6 & 1.9 & 9.81 & 3.03 & -1.09 \\
163 & 4FGL J2040.1-4621  & 24.6 & 4.7 & FSPT-S J204006-4620.3 & 4.6 & 0.9 & 1.69 & 1.18 & -2.48 \\
164 & 4FGL J0437.4-6155  & 24.5 & 3.6 & SPT-S J043718-6157.0 & 5.9 & 1.6 & 1.33 & 0.61 & -2.87 \\
165 & 4FGL J0026.6-4600  & 24.3 & 3.9 & SPT-S J002636-4601.1 & 8.9 & 2.2 & 0.19 & 0.18 & -4.65 \\
166 & 4FGL J0009.8-4317  & 24.3 & 3.7 & SPT-S J000949-4316.7 & 18.2 & 3.8 & 1.14 & 0.99 & -3.35 \\
167 & 4FGL J0017.1-4605* & 24.2 & 4.1 & - & \multicolumn{1}{r}{-} &  & - & - & - \\
168 & 4FGL J2105.2-5143  & 24.2 & 4.7 & SPT-S J210524-5145.7 & 26.6 & 5.5 & 2.36 & 0.93 & -2.88 \\
169 & 4FGL J0225.5-5530* & 24.1 & 11.7 & FSPT-S J022532-5528.6 & 4.0 & 0.8 & 1.68 & 0.86 & -2.50 \\
170 & 4FGL J0654.6-4952  & 24.1 & 4.6 & SPT-S J065519-4951.9 & 38.5 & 7.8 & 6.30 & 1.56 & -2.20 \\
171 & 4FGL J0019.2-5640  & 24.1 & 3.5 & SPT-S J001926-5641.7 & 79.4 & 16.0 & 1.85 & 0.97 & -3.67 \\
172 & 4FGL J2353.1-4806  & 23.9 & 3.7 & SPT-S J235311-4806.0 & 62.0 & 12.5 & 0.68 & 0.56 & -4.43 \\
173 & 4FGL J0550.5-4356* & 23.9 & 4.3 & FSPT-S J055026-4356.9 & 4.8 & 1.0 & 0.87 & 0.77 & -2.99 \\
174 & 4FGL J2209.8-5028  & 23.8 & 4.5 & SPT-S J221040-5026.9 & 98.5 & 19.8 & 7.55 & 2.02 & -2.66 \\
& & \multicolumn{1}{r}{} & & SPT-S J221015-5031.1 & 25.4 & 5.3 & 4.58 & 1.33 & -2.40 \\
175 & 4FGL J0357.6-4625  & 23.4 & 3.5 & SPT-S J035728-4625.6 & 69.1 & 13.9 & 1.98 & 1.08 & -3.51 \\
176 & 4FGL J0533.3-5549  & 23.3 & 4.8 & SPT-S J053324-5549.5 & 57.8 & 11.6 & 0.76 & 0.10 & -4.38 \\
177 & 4FGL J0343.3-6303* & 23.2 & 3.8 & SPT-S J034325-6303.3 & 8.4 & 2.0 & 0.30 & 0.20 & -4.23 \\
178 & 4FGL J0131.7-5346* & 23.2 & 3.8 & - & \multicolumn{1}{r}{-} &  & - & - & - \\
179 & 4FGL J0140.5-4730* & 23.1 & 4.5 & SPT-S J013940-4732.0 & 48.8 & 9.9 & 8.94 & 3.13 & -2.08 \\
& & \multicolumn{1}{r}{} & & SPT-S J014046-4725.7 & 5.9 & 1.6 & 5.12 & 1.83 & -1.54 \\
180 & 4FGL J0009.1-5012* & 23.0 & 4.0 & SPT-S J000835-5009.6 & 13.7 & 3.0 & 6.36 & 1.46 & -1.80 \\
181 & 4FGL J2249.7-5944  & 22.9 & 3.8 & SPT-S J224938-5944.2 & 6.2 & 1.7 & 0.69 & 0.54 & -3.36 \\
182 & 4FGL J0624.7-4903* & 22.9 & 4.5 & FSPT-S J062358-4904.1 & 3.6 & 0.7 & 8.08 & 1.76 & -1.15 \\
183 & 4FGL J0110.0-4019  & 22.8 & 4.3 & SPT-S J010956-4020.7 & 12.3 & 2.8 & 2.23 & 2.01 & -2.64 \\
184 & 4FGL J2029.5-4237* & 22.7 & 2.8 & - & \multicolumn{1}{r}{-} &  & - & - & - \\
185 & 4FGL J0035.0-5728  & 22.7 & 4.0 & SPT-S J003504-5726.2 & 17.5 & 3.7 & 2.07 & 0.47 & -2.95 \\
186 & 4FGL J0225.6-4502  & 22.5 & 5.4 & SPT-S J022544-4503.2 & 355.8 & 71.3 & 1.65 & 0.74 & -4.46 \\
187 & 4FGL J2107.6-4148  & 22.5 & 4.5 & SPT-S J210723-4145.5 & 23.8 & 4.9 & 4.39 & 1.26 & -2.39 \\
188 & 4FGL J0102.6-5639  & 22.5 & 4.5 & SPT-S J010210-5637.2 & 120.3 & 24.1 & 4.32 & 1.94 & -3.12 \\
& & \multicolumn{1}{r}{} & & SPT-S J010303-5639.3 & 5.9 & 1.7 & 3.57 & 1.77 & -1.75 \\
189 & 4FGL J0049.6-4500  & 22.3 & 3.8 & SPT-S J004916-4457.1 & 1158.0 & 232.0 & 5.35 & 2.23 & -4.34 \\
190 & 4FGL J0358.1-5954  & 22.2 & 3.5 & SPT-S J035814-5952.3 & 21.4 & 4.4 & 2.18 & 0.85 & -2.90 \\
191 & 4FGL J0056.6-5317  & 22.1 & 3.5 & SPT-S J005621-5318.6 & 20.3 & 4.3 & 2.65 & 1.75 & -2.70 \\
192 & 4FGL J0440.3-4333  & 22.0 & 5.5 & SPT-S J044017-4333.0 & 161.5 & 32.4 & 0.77 & 0.15 & -4.61 \\
& & \multicolumn{1}{r}{} & & SPT-S J044117-4313.6 & 111.3 & 22.3 & 22.05 & 2.97 & -1.81 \\
193 & 4FGL J0623.9-5259  & 22.0 & 4.1 & SPT-S J062337-5258.3 & 7.9 & 2.0 & 3.07 & 0.78 & -2.21 \\
& & \multicolumn{1}{r}{} & & SPT-S J062337-5257.8 & 3.3 & 1.3 & 3.32 & 0.82 & -1.89 \\
194 & 4FGL J0414.7-4300* & 22.0 & 2.7 & - & \multicolumn{1}{r}{-} &  & - & - & - \\
195 & 4FGL J0604.5-4851  & 21.8 & 4.7 & SPT-S J060433-4849.6 & 155.4 & 31.2 & 1.83 & 1.22 & -3.95 \\
196 & 4FGL J0647.0-5138  & 21.8 & 4.3 & SPT-S J064709-5135.8 & 9.4 & 2.3 & 2.54 & 1.83 & -2.44 \\
197 & 4FGL J0004.4-4001* & 21.8 & 4.4 & SPT-S J000433-4000.5 & 19.1 & 4.0 & 1.33 & 0.49 & -3.13 \\
& & \multicolumn{1}{r}{} & & SPT-S J000444-4007.3 & 0.7 & 1.5 & 6.58 & 1.85 & -1.25 \\
198 & 4FGL J0601.4-6057* & 21.6 & 4.4 & - & \multicolumn{1}{r}{-} &  & - & - & - \\
199 & 4FGL J0500.6-4911  & 21.5 & 4.0 & SPT-S J050037-4912.1 & 15.5 & 3.3 & 1.22 & 1.21 & -3.25 \\
200 & 4FGL J0658.1-5840  & 21.2 & 4.1 & SPT-S J065814-5840.3 & 123.9 & 24.9 & 0.69 & 0.40 & -4.88 \\
201 & 4FGL J2357.0-4840  & 21.1 & 3.7 & SPT-S J235721-4838.3 & 111.7 & 22.4 & 3.51 & 2.19 & -3.30 \\
202 & 4FGL J2247.7-5857* & 20.8 & 4.5 & FSPT-S J224745-5854.9 & 4.1 & 0.8 & 2.44 & 0.70 & -2.14 \\
203 & 4FGL J0127.4-4813  & 20.6 & 3.4 & SPT-S J012715-4813.4 & 127.5 & 25.6 & 2.27 & 1.07 & -3.77 \\
204 & 4FGL J0025.4-4838* & 20.5 & 3.9 & - & \multicolumn{1}{r}{-} &  & - & - & - \\
205 & 4FGL J2031.2-4121  & 20.5 & 4.2 & SPT-S J203055-4117.1 & 35.9 & 7.3 & 5.63 & 1.90 & -2.22 \\
206 & 4FGL J0438.2-4243* & 20.4 & 4.4 & SPT-S J043831-4240.0 & 5.8 & 1.6 & 4.72 & 2.15 & -1.60 \\
207 & 4FGL J2246.7-5207  & 20.4 & 4.3 & FSPT-S J224642-5206.6 & 4.4 & 0.9 & 0.91 & 0.89 & -2.95 \\
208 & 4FGL J0245.4-5950  & 20.4 & 3.7 & SPT-S J024452-5947.9 & 29.3 & 6.0 & 4.49 & 1.98 & -2.36 \\
209 & 4FGL J2151.2-4034* & 20.3 & 4.1 & - & \multicolumn{1}{r}{-} &  & - & - & - \\
210 & 4FGL J2240.7-4746  & 20.1 & 3.6 & SPT-S J224043-4747.3 & 10.1 & 2.4 & 0.96 & 0.96 & -3.31 \\
211 & 4FGL J0047.1-6203  & 20.1 & 4.1 & SPT-S J004750-6206.7 & 5.5 & 1.5 & 6.30 & 1.16 & -1.27 \\
212 & 4FGL J0622.7-4141  & 20.0 & 3.2 & - & \multicolumn{1}{r}{-} &  & - & - & - \\
213 & 4FGL J0119.4-5354  & 19.8 & 3.6 & SPT-S J011950-5357.2 & 189.4 & 38.0 & 3.98 & 1.36 & -3.46 \\
214 & 4FGL J2332.1-4118  & 19.8 & 4.5 & SPT-S J233218-4118.6 & 113.7 & 22.8 & 2.44 & 1.10 & -3.54 \\
215 & 4FGL J0541.1-4854  & 19.5 & 3.9 & - & \multicolumn{1}{r}{-} &  & - & - & - \\
216 & 4FGL J0553.9-5048* & 19.4 & 3.7 & FSPT-S J055359-5051.7 & 3.4 & 0.7 & 2.97 & 1.64 & -1.93 \\
217 & 4FGL J0650.2-5144* & 19.3 & 4.1 & SPT-S J065009-5144.5 & 7.7 & 2.0 & 1.13 & 1.03 & -3.04 \\
218 & 4FGL J0206.8-5744  & 19.2 & 3.4 & SPT-S J020641-5749.7 & 6.4 & 1.7 & 4.97 & 1.66 & -1.82 \\
& & \multicolumn{1}{r}{} & & SPT-S J020721-5751.3 & 3.5 & 1.2 & 7.73 & 3.05 & -1.11 \\
219 & 4FGL J0420.3-6016  & 19.2 & 3.3 & - & \multicolumn{1}{r}{-} &  & - & - & - \\
220 & 4FGL J0159.3-4523  & 19.2 & 5.5 & SPT-S J015906-4515.6 & 42.2 & 8.6 & 7.69 & 1.67 & -2.18 \\
221 & 4FGL J0550.3-5733  & 19.0 & 4.3 & SPT-S J055009-5732.4 & 542.9 & 108.8 & 1.68 & 0.88 & -4.80 \\
222 & 4FGL J2355.2-5247* & 18.8 & 4.0 & - & \multicolumn{1}{r}{-} &  & - & - & - \\
223 & 4FGL J0414.8-5338  & 18.6 & 4.4 & FSPT-S J041458-5339.7 & 4.0 & 0.8 & 1.43 & 1.07 & -2.60 \\
224 & 4FGL J0406.0-5407  & 18.5 & 3.7 & SPT-S J040608-5404.7 & 8.6 & 2.0 & 2.73 & 1.09 & -2.31 \\
225 & 4FGL J2321.9-4842* & 18.3 & 4.2 & SPT-S J232216-4836.2 & 7.7 & 2.0 & 6.83 & 3.29 & -1.25 \\
226 & 4FGL J2012.1-5234* & 17.9 & 4.3 & SPT-S J201213-5232.8 & 7.3 & 1.9 & 1.54 & 1.24 & -2.64 \\
& & \multicolumn{1}{r}{} & & SPT-S J201142-5235.2 & 24.8 & 5.1 & 4.16 & 2.94 & -2.46 \\
227 & 4FGL J2213.5-4754  & 17.9 & 3.8 & SPT-S J221330-4754.4 & 8.5 & 2.1 & 0.44 & 0.38 & -3.87 \\
228 & 4FGL J0214.8-6150  & 17.7 & 3.8 & SPT-S J021415-6149.5 & 385.3 & 77.2 & 3.99 & 1.02 & -3.82 \\
229 & 4FGL J0350.4-5144  & 17.0 & 3.5 & FSPT-S J035028-5144.7 & 4.1 & 0.8 & 0.20 & 0.15 & -4.27 \\
230 & 4FGL J0150.6-5448  & 17.0 & 3.3 & SPT-S J015044-5450.1 & 10.2 & 2.3 & 2.02 & 1.25 & -2.67 \\
231 & 4FGL J0443.4-4152  & 17.0 & 3.6 & SPT-S J044328-4151.6 & 6.1 & 1.7 & 1.41 & 0.81 & -2.69 \\
232 & 4FGL J2046.9-5409* & 16.9 & 4.0 & FSPT-S J204701-5412.7 & 3.8 & 0.8 & 2.96 & 1.69 & -1.94 \\
233 & 4FGL J0003.3-5905  & 16.8 & 3.4 & SPT-S J000312-5905.7 & 12.6 & 2.7 & 1.37 & 0.56 & -2.92 \\
234 & 4FGL J0125.9-6303* & 16.6 & 3.4 & FSPT-S J012541-6305.7 & 2.2 & 0.4 & 3.11 & 1.42 & -1.86 \\
& & \multicolumn{1}{r}{} & & FSPT-S J012547-6302.7 & 6.1 & 1.2 & 1.53 & 0.74 & -2.61 \\
235 & 4FGL J0539.2-6333* & 16.6 & 3.7 & - & \multicolumn{1}{r}{-} &  & - & - & - \\
236 & 4FGL J0654.0-4152  & 16.4 & 4.5 & SPT-S J065400-4151.8 & 28.6 & 5.9 & 1.25 & 0.55 & -3.47 \\
237 & 4FGL J0301.6-5617* & 16.2 & 3.3 & FSPT-S J030115-5616.7 & 3.5 & 0.7 & 3.57 & 1.83 & -1.76 \\
238 & 4FGL J0652.1-4813  & 16.0 & 3.9 & SPT-S J065203-4809.0 & 6.2 & 1.7 & 4.33 & 1.61 & -1.77 \\
239 & 4FGL J2030.3-5038* & 16.0 & 3.9 & - & \multicolumn{1}{r}{-} &  & - & - & - \\
240 & 4FGL J0056.6-4452  & 15.8 & 3.6 & SPT-S J005646-4451.0 & 91.8 & 18.4 & 2.10 & 0.53 & -3.75 \\
241 & 4FGL J0407.7-5702* & 15.7 & 3.6 & - & \multicolumn{1}{r}{-} &  & - & - & - \\
242 & 4FGL J0314.4-4805* & 15.7 & 3.3 & SPT-S J031428-4807.8 & 10.1 & 2.4 & 2.76 & 0.69 & -2.33 \\
243 & 4FGL J0031.5-5648* & 15.4 & 3.6 & FSPT-S J003136-5646.6 & 2.3 & 0.5 & 1.80 & 0.58 & -2.35 \\
244 & 4FGL J0429.3-4326  & 15.3 & 3.5 & SPT-S J042924-4328.5 & 15.2 & 3.3 & 1.98 & 0.83 & -2.87 \\
245 & 4FGL J0309.4-4000  & 15.0 & 3.4 & SPT-S J030912-4001.8 & 27.8 & 5.7 & 3.07 & 2.04 & -2.68 \\
246 & 4FGL J0049.5-4150  & 14.8 & 3.6 & SPT-S J004939-4151.3 & 4.0 & 1.6 & 1.06 & 0.58 & -2.82 \\
247 & 4FGL J2343.0-4756* & 14.8 & 2.7 & FSPT-S J234302-4757.8 & 5.5 & 1.1 & 1.07 & 0.53 & -2.90 \\
248 & 4FGL J0610.8-4911* & 14.7 & 3.8 & - & \multicolumn{1}{r}{-} &  & - & - & - \\
249 & 4FGL J2311.6-4427* & 14.6 & 3.8 & - & \multicolumn{1}{r}{-} &  & - & - & - \\
250 & 4FGL J0102.0-6240* & 14.4 & 3.3 & FSPT-S J010147-6243.1 & 3.9 & 0.8 & 2.73 & 1.08 & -2.05 \\
251 & 4FGL J0611.4-4722* & 14.3 & 3.7 & - & \multicolumn{1}{r}{-} &  & - & - & - \\
252 & 4FGL J2042.1-5320* & 14.2 & 3.9 & FSPT-S J204217-5321.1 & 2.3 & 0.5 & 1.62 & 0.62 & -2.28 \\
& & \multicolumn{1}{r}{} & & FSPT-S J204220-5326.9 & 4.7 & 0.9 & 6.96 & 1.92 & -1.39 \\
253 & 4FGL J0401.0-5353  & 14.1 & 3.6 & - & \multicolumn{1}{r}{-} &  & - & - & - \\
254 & 4FGL J2127.6-5959  & 13.7 & 3.2 & SPT-S J212722-6000.8 & 12.9 & 2.8 & 2.80 & 1.54 & -2.38 \\
255 & 4FGL J2316.9-5210  & 13.6 & 3.3 & SPT-S J231702-5210.0 & 8.2 & 1.9 & 0.55 & 0.32 & -3.55 \\
256 & 4FGL J2042.7-5415  & 13.6 & 2.2 & FSPT-S J204304-5411.6 & 2.1 & 0.4 & 4.89 & 1.76 & -1.44 \\
257 & 4FGL J2239.2-5657  & 13.6 & 3.3 & SPT-S J223911-5701.0 & 520.7 & 104.3 & 3.95 & 1.00 & -3.85 \\
258 & 4FGL J2202.7-5637  & 13.4 & 3.4 & SPT-S J220253-5635.7 & 40.5 & 8.2 & 2.30 & 1.38 & -3.25 \\
259 & 4FGL J0246.0-4838* & 13.4 & 4.1 & - & \multicolumn{1}{r}{-} &  & - & - & - \\
260 & 4FGL J0459.7-5413* & 13.3 & 3.2 & - & \multicolumn{1}{r}{-} &  & - & - & - \\
261 & 4FGL J0023.6-4209* & 13.3 & 3.3 & SPT-S J002443-4202.2 & 12.6 & 2.8 & 14.13 & 3.25 & -1.32 \\
& & \multicolumn{1}{r}{} & & SPT-S J002401-4200.9 & 8.6 & 2.1 & 9.63 & 2.79 & -1.27 \\
& & \multicolumn{1}{r}{} & & SPT-S J002300-4206.4 & 8.0 & 2.0 & 7.63 & 2.69 & -1.27 \\
262 & 4FGL J0311.5-4402  & 13.2 & 3.2 & FSPT-S J031103-4402.5 & 1.8 & 0.4 & 5.08 & 1.51 & -1.53 \\
263 & 4FGL J0214.4-5822  & 12.9 & 3.1 & SPT-S J021409-5822.0 & 46.5 & 9.4 & 1.95 & 0.82 & -3.42 \\
264 & 4FGL J0450.7-4938  & 12.8 & 3.2 & SPT-S J045101-4936.3 & 80.8 & 16.2 & 3.76 & 1.13 & -3.16 \\
265 & 4FGL J0331.1-5243  & 12.4 & 3.1 & SPT-S J033113-5241.7 & 27.2 & 5.6 & 2.27 & 0.53 & -3.18 \\
& & \multicolumn{1}{r}{} & & SPT-S J033124-5258.4 & 118.9 & 23.8 & 14.64 & 3.64 & -2.21 \\
266 & 4FGL J2321.0-6308  & 12.3 & 3.0 & SPT-S J232042-6309.7 & 3.0 & 1.5 & 2.62 & 2.07 & -1.99 \\
267 & 4FGL J0328.8-5715  & 12.1 & 3.4 & - & \multicolumn{1}{r}{-} &  & - & - & - \\
268 & 4FGL J0316.0-5626  & 11.8 & 2.9 & - & \multicolumn{1}{r}{-} &  & - & - & - \\
269 & 4FGL J0058.3-4603* & 11.4 & 3.1 & - & \multicolumn{1}{r}{-} &  & - & - & - \\
270 & 4FGL J0640.9-5204* & 11.4 & 3.3 & SPT-S J064110-5202.5 & 24.4 & 5.1 & 2.62 & 1.67 & -2.81 \\
271 & 4FGL J0106.9-4832  & 11.0 & 2.7 & SPT-S J010655-4831.4 & 27.3 & 5.6 & 0.78 & 0.25 & -4.00 \\
272 & 4FGL J0118.3-6008* & 11.0 & 2.8 & SPT-S J011823-6007.8 & 10.8 & 2.5 & 0.82 & 0.25 & -3.43 \\
273 & 4FGL J0347.0-6400* & 9.8 & 1.8 & - & \multicolumn{1}{r}{-} &  & - & - & - \\
274 & 4FGL J0133.2-4533  & 9.7 & 3.0 & FSPT-S J013309-4535.4 & 1.5 & 0.3 & 1.95 & 1.36 & -2.17 \\
275 & 4FGL J0215.0-5330* & 9.7 & 2.9 & FSPT-S J021515-5328.7 & 4.3 & 0.9 & 2.30 & 1.15 & -2.16 \\
276 & 4FGL J0338.7-5706  & 9.4 & 3.0 & - & \multicolumn{1}{r}{-} &  & - & - & - \\
277 & 4FGL J0101.7-5455  & 9.3 & 2.9 & FSPT-S J010141-5455.8 & 4.7 & 0.9 & 0.79 & 0.61 & -3.12 \\
278 & 4FGL J0620.7-5034* & 9.2 & 3.0 & SPT-S J062045-5033.9 & 8.0 & 2.0 & 0.47 & 0.27 & -3.81 \\
279 & 4FGL J0317.8-4414  & 8.5 & 3.1 & SPT-S J031757-4414.0 & 74.7 & 15.0 & 0.82 & 0.42 & -4.42 \\
280 & 4FGL J0201.1-4347  & 8.1 & 3.0 & FSPT-S J020110-4346.8 & 4.4 & 0.9 & 1.00 & 0.75 & -2.91 \\
281 & 4FGL J0308.9-4702* & 6.9 & 2.7 & FSPT-S J030858-4700.5 & 1.5 & 0.3 & 1.92 & 1.45 & -2.25 \\

\end{longtable}
\end{center}

\section{B. Full 4FGL Multi-wavelength Association Table}
The full 4FGL multi-wavelength association table is available online\footnote{https://github.com/lizhong-phys/4FGL-SPT.git}. Note the index and table format are consistent with \autoref{tab:spt_4fgl}. Additionally, the source name of multi-wavelength counterpart (SUMSS/SPT/\textit{WISE}/RASS) starting with `*' represents that the source satisfies the selection criteria described in the \textit{Method} \autoref{sec:method}. 
For each association, multiple counterparts are ranked based on the \pvalue from low to high. We attach an example table for illustration. The column description and units are listed below.
\\

1. 4FGL Name: 4FGL source name \\

2. SPT Name: SPT source name \\

3. SPT RA: SPT right ascension (J2000) in degrees \\

4. SPT DEC: SPT declination (J2000) in degrees \\

5. $S_{95}$: (Deboosted) flux in 95GHz in mJy \\

6. $S_{150}$: (Deboosted) flux in 150GHz in mJy \\

7. $S_{220}$: (Deboosted) flux in 220GHz in mJy \\

8. Redshift: Measured redshift if available in NED \\

9. RASS name: RASS source name \\

10. SUMSS name: SUMSS source name \\

11. WISE name: AllWISE source name \\

\begin{table*}[htbp]
\centering
\caption[]{Example Table of Full 4FGL Multi-Wavelength Association} 
\tiny
\renewcommand{\arraystretch}{1.5}
\begin{tabular}{ l c c c c c c c c c c c c }
\hline\hline
Index & 4FGL Name & SPT Name & SPT R.A. & SPT decl. & $S_{\mathrm{95GHz}}$ & $S_{\mathrm{150GHz}}$ & $S_{\mathrm{220GHz}}$ & z & RASS Name & SUMSS Name & AllWISE Name \\
& (4FGL *) & (SPT-S *) & (deg) & (deg) & (mJy) & (mJy) & (mJy) & & (RASS *) & (SUMSS *) & (AllWISE *) \\
\hline
0 & J0538.8-4405 & J053850-4405.1 & 84.710495 & -44.085197 & 5958.9 & 4843.0 & 3905.6 & 0.89400 & J053850.2-440504 & J053850-440510 & J053850.36-440508.9 \\
  &              & -              & -         & -          & -      & -      & -      & -       & -                & J053849-440414 & -                   \\
1 & J2329.3-4955 & J232920-4955.6 & 352.33704 & -49.927555 & 1064.1 & 522.1  & 426.4  & 0.51800 & -                & J232920-495540 & J232920.87-495540.6 \\
2 & J0449.4-4350 & J044924-4350.0 & 72.351730 & -43.833607 & 127.9  & 119.3  & 108.7  & 0.10700 & J044924.2-435002 & J044924-435011 & J044924.69-435008.9 \\
3 & J2056.2-4714 & J205616-4714.8 & 314.06787 & -47.247147 & 1769.5 & 1424.9 & 1199.5 & 1.48900 & J205615.8-471446 & J205616-471448 & J205616.36-471447.6 \\
4 & J0210.7-5101 & J021045-5101.0 & 32.689522 & -51.017540 & 1868.2 & 1504.4 & 1196.5 & 0.99900 & J021046.8-510055 & J021046-510102 & J021046.20-510101.8 \\
  &              & -              & -         & -          & -      & -      & -      & -       & -                & J021055-510017 & -                   \\
5 & J0532.0-4827 & J053158-4827.6 & 82.994888 & -48.461178 & 450.9  & 400.3  & 353.0  & -       & J053159.9-482751 & J053158-482737 & J053158.62-482735.9 \\
\hline
\end{tabular}
\label{tab:sample}
\footnotesize
\tablecomments{Full table in multiple file formats can be found in \url{https://github.com/lizhong-phys/4FGL-SPT.git}.}
\end{table*}

\section{C. Thumbnails} 
\label{sec:thumbnails}
We make the thumbnails of all 282 4FGL sources within 2500 $\mathrm{deg^2}$ SPT-SZ survey field, similar to \autoref{fig:ex_method}. The index is consistent with \autoref{tab:spt_4fgl} where the 282 4FGL sources are sorted by $\gamma$-ray flux. \lz{The thumbnails are attached below and can also be downloaded via \url{https://github.com/lizhong-phys/4FGL-SPT.git}.}  Each $0.7^{\circ} \times 0.7^{\circ}$ thumbnail in the grey background is the high-pass filtered SPT 150 GHz image. The SPT images appear to be ringing because of the high-pass point source filter. The blue ellipse at the center is the 4FGL 95\% ($2\sigma$) uncertainty region and the blue dashed ellipse represents the $4\sigma$ 4FGL beam. The green diamond marks the position of SPT-SZ sources at $>4.5\sigma$ from \citet{2020arXiv200303431E}. The red contours are derived from the 843 MHz SUMSS map in units of 3,4,5,7,10,15,20,25$\sigma$. Each radio detection in the SUMSS catalog is also marked by a red dot for the cases too faint to be seen in contours. The 4FGL source name is at the upper-left corner as well as the $\gamma$-ray energy flux. 

\figsetstart
\figsetnum{12}
\figsettitle{SPT thumbnails of 282 4FGL sources}

\figsetgrpstart
\figsetgrpnum{12.1}
\figsetgrptitle{1}
\figsetplot{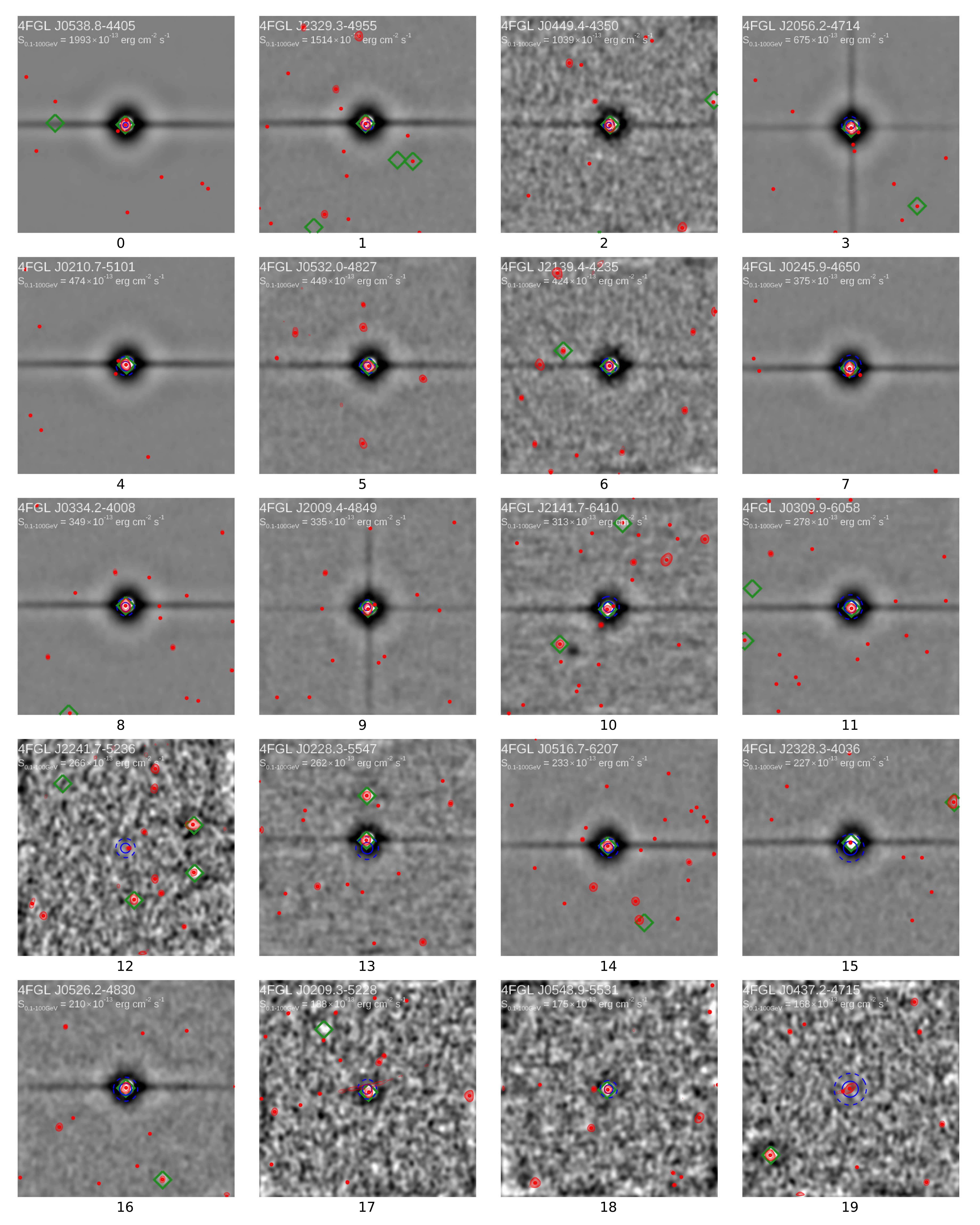}
\figsetgrpnote{SPT thumbnails of 282 4FGL sources. The full version can also be downloaded in \url{https://github.com/lizhong-phys/4FGL-SPT.git}.
}
\figsetgrpend

\figsetgrpstart
\figsetgrpnum{12.2}
\figsetgrptitle{2}
\figsetplot{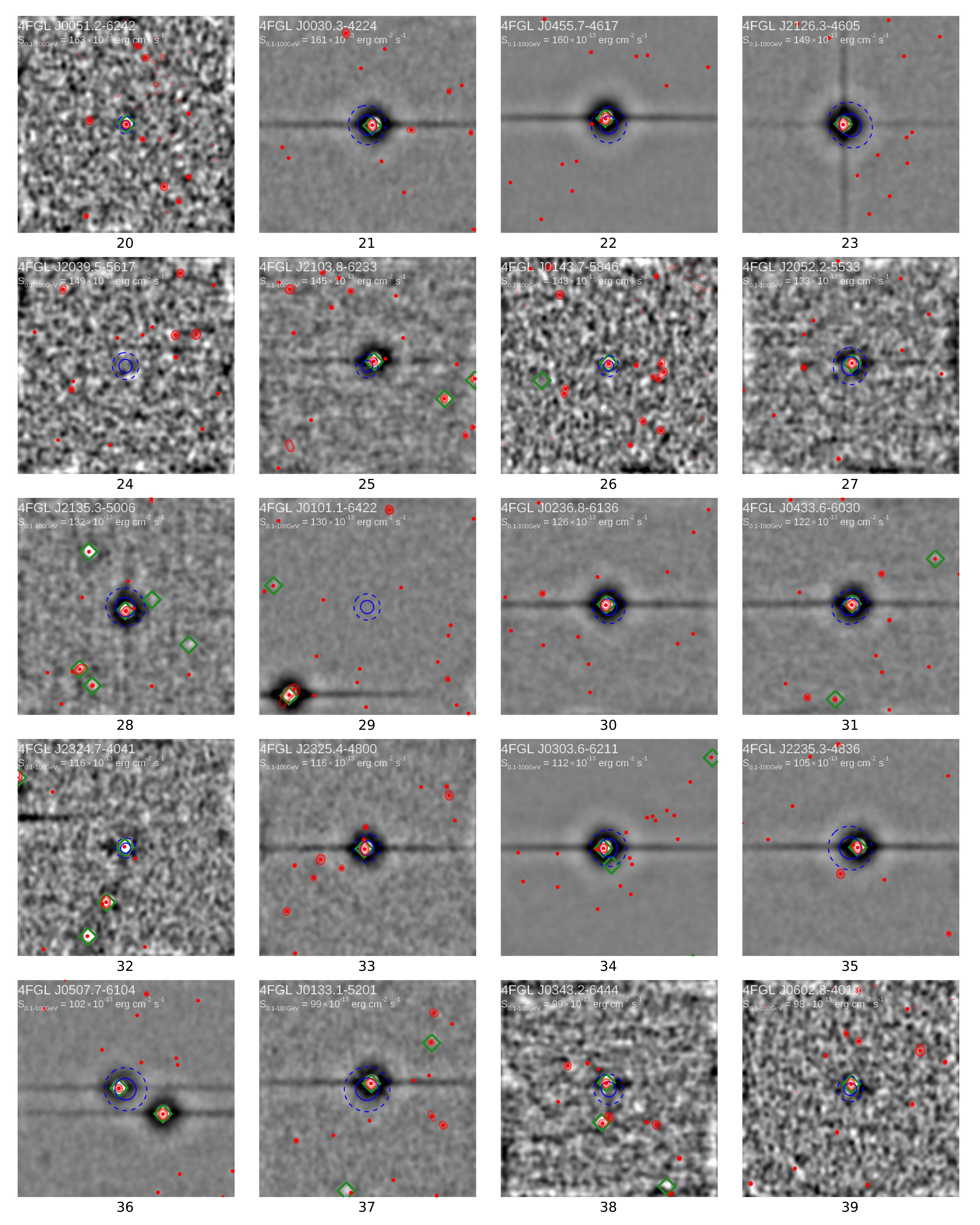}
\figsetgrpnote{SPT thumbnails of 282 4FGL sources. The full version can also be downloaded in \url{https://github.com/lizhong-phys/4FGL-SPT.git}.
}
\figsetgrpend

\figsetgrpstart
\figsetgrpnum{12.3}
\figsetgrptitle{3}
\figsetplot{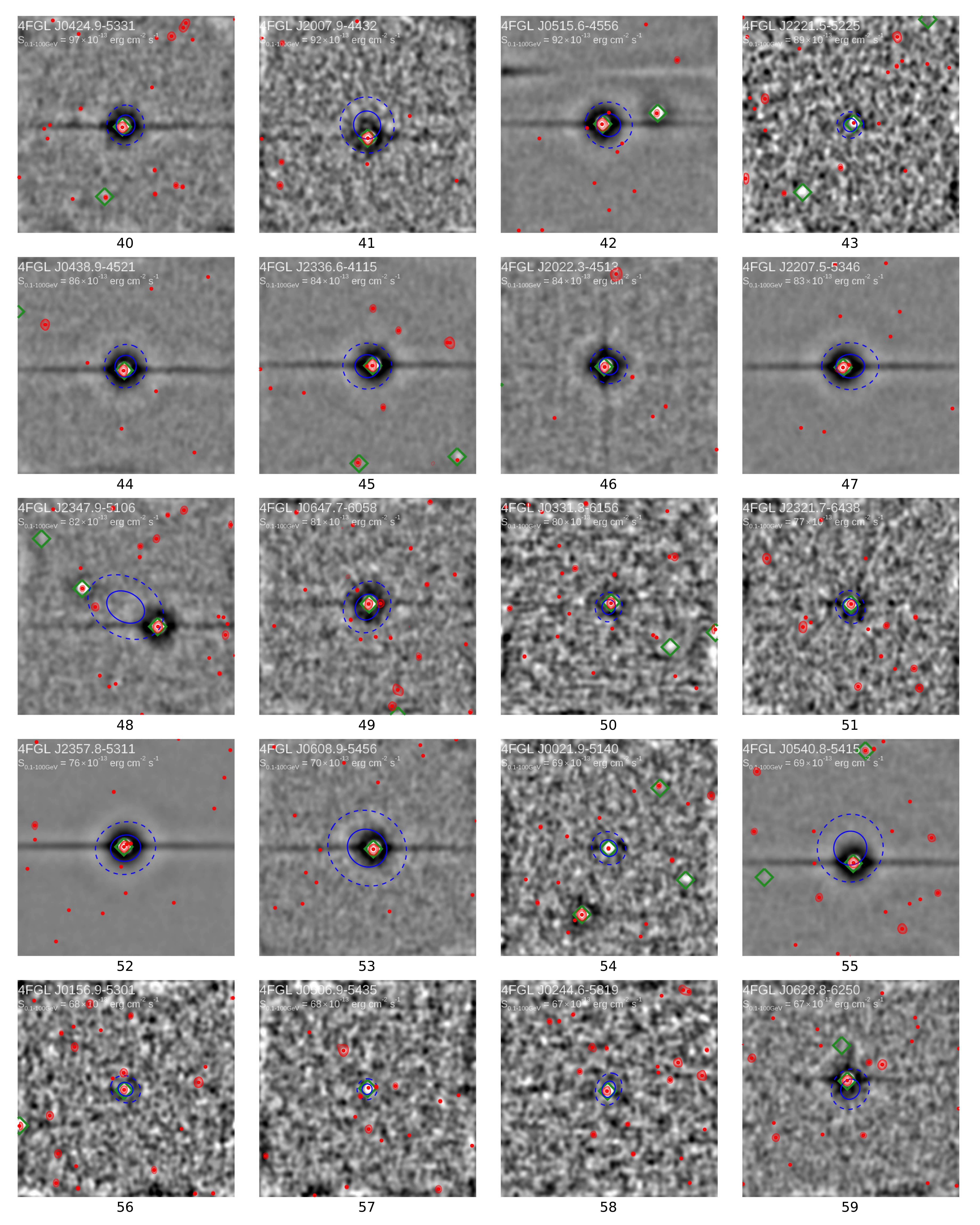}
\figsetgrpnote{SPT thumbnails of 282 4FGL sources. The full version can also be downloaded in \url{https://github.com/lizhong-phys/4FGL-SPT.git}.
}
\figsetgrpend

\figsetgrpstart
\figsetgrpnum{12.4}
\figsetgrptitle{4}
\figsetplot{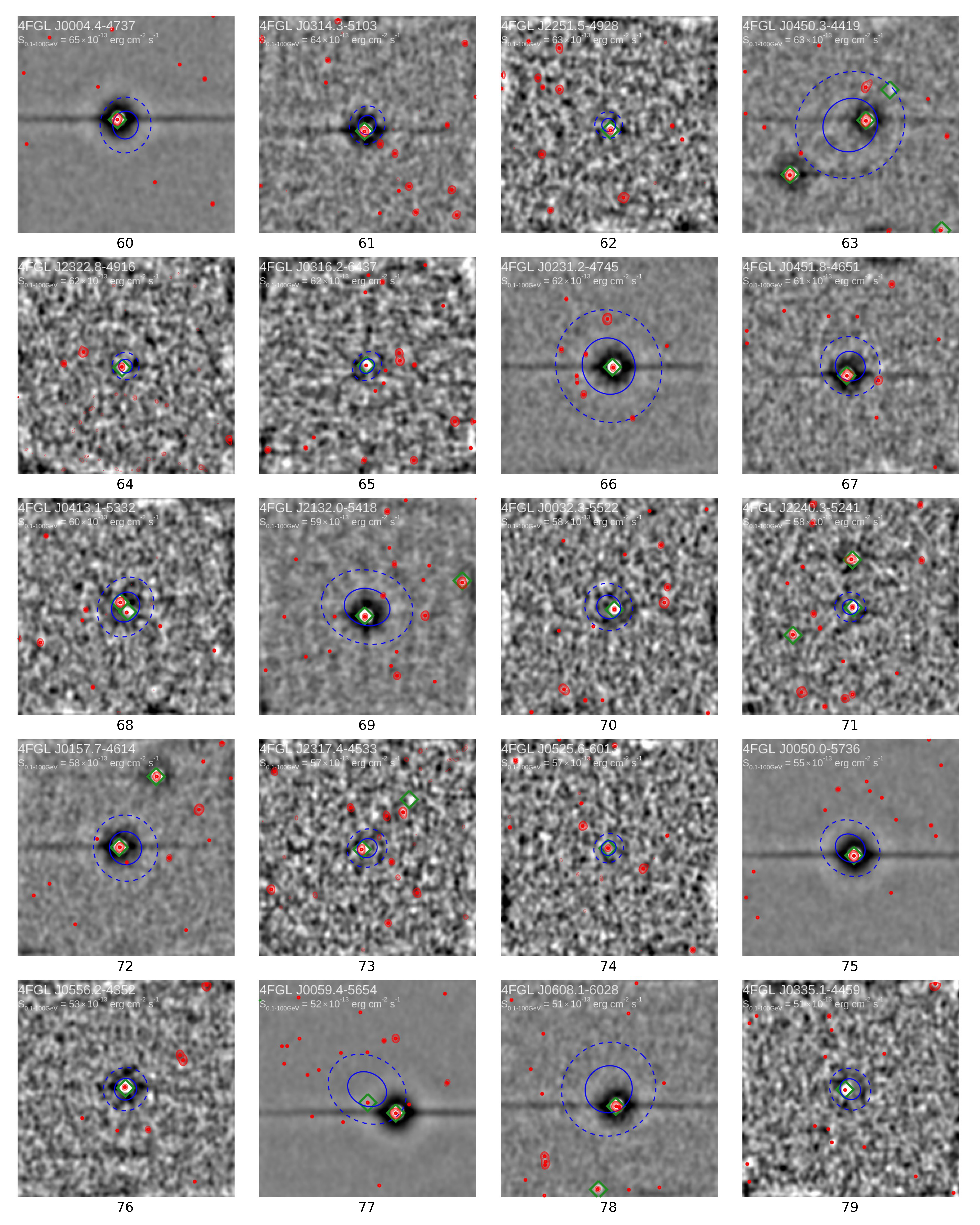}
\figsetgrpnote{SPT thumbnails of 282 4FGL sources. The full version can also be downloaded in \url{https://github.com/lizhong-phys/4FGL-SPT.git}.
}
\figsetgrpend

\figsetgrpstart
\figsetgrpnum{12.5}
\figsetgrptitle{5}
\figsetplot{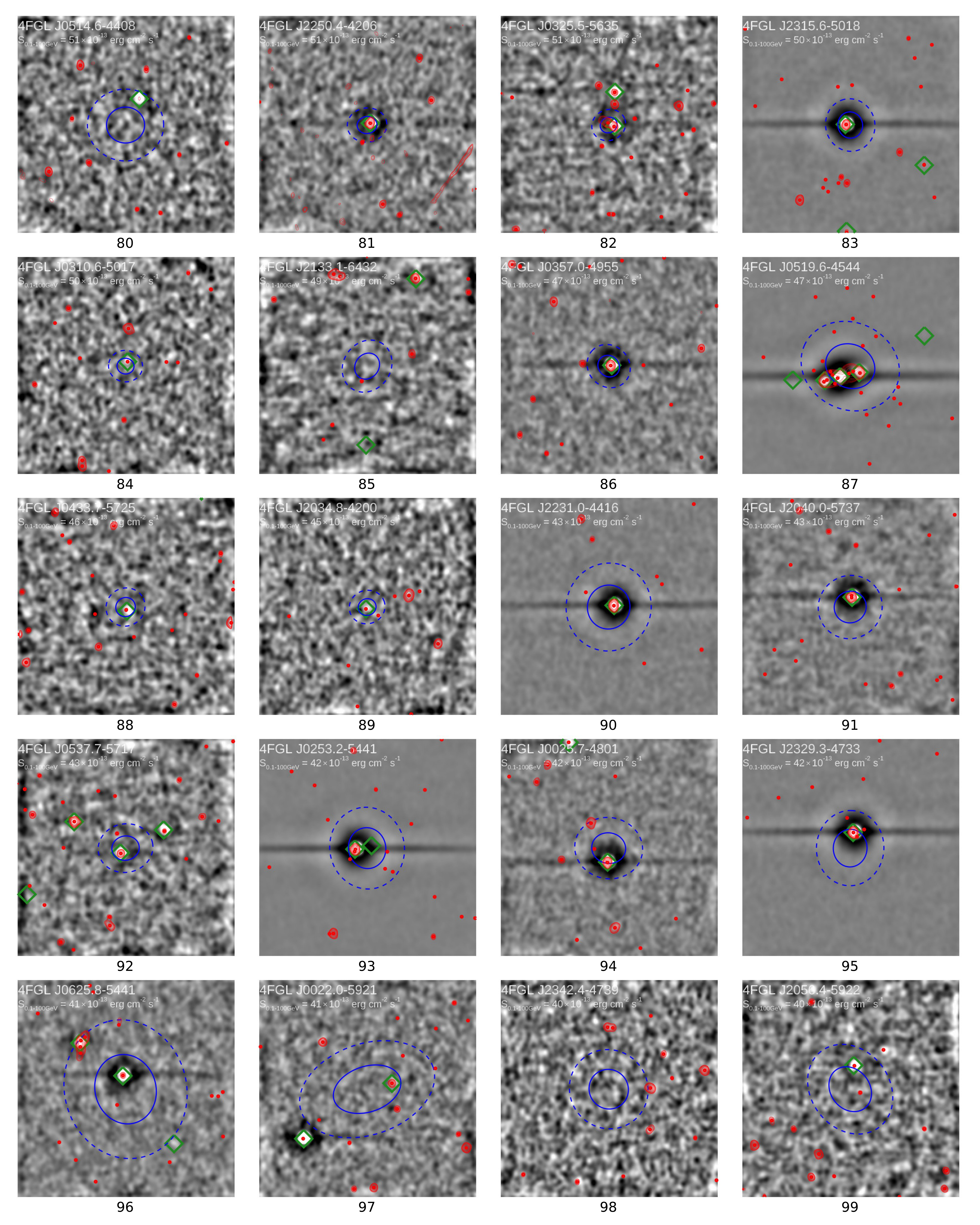}
\figsetgrpnote{SPT thumbnails of 282 4FGL sources. The full version can also be downloaded in \url{https://github.com/lizhong-phys/4FGL-SPT.git}.
}
\figsetgrpend

\figsetgrpstart
\figsetgrpnum{12.6}
\figsetgrptitle{6}
\figsetplot{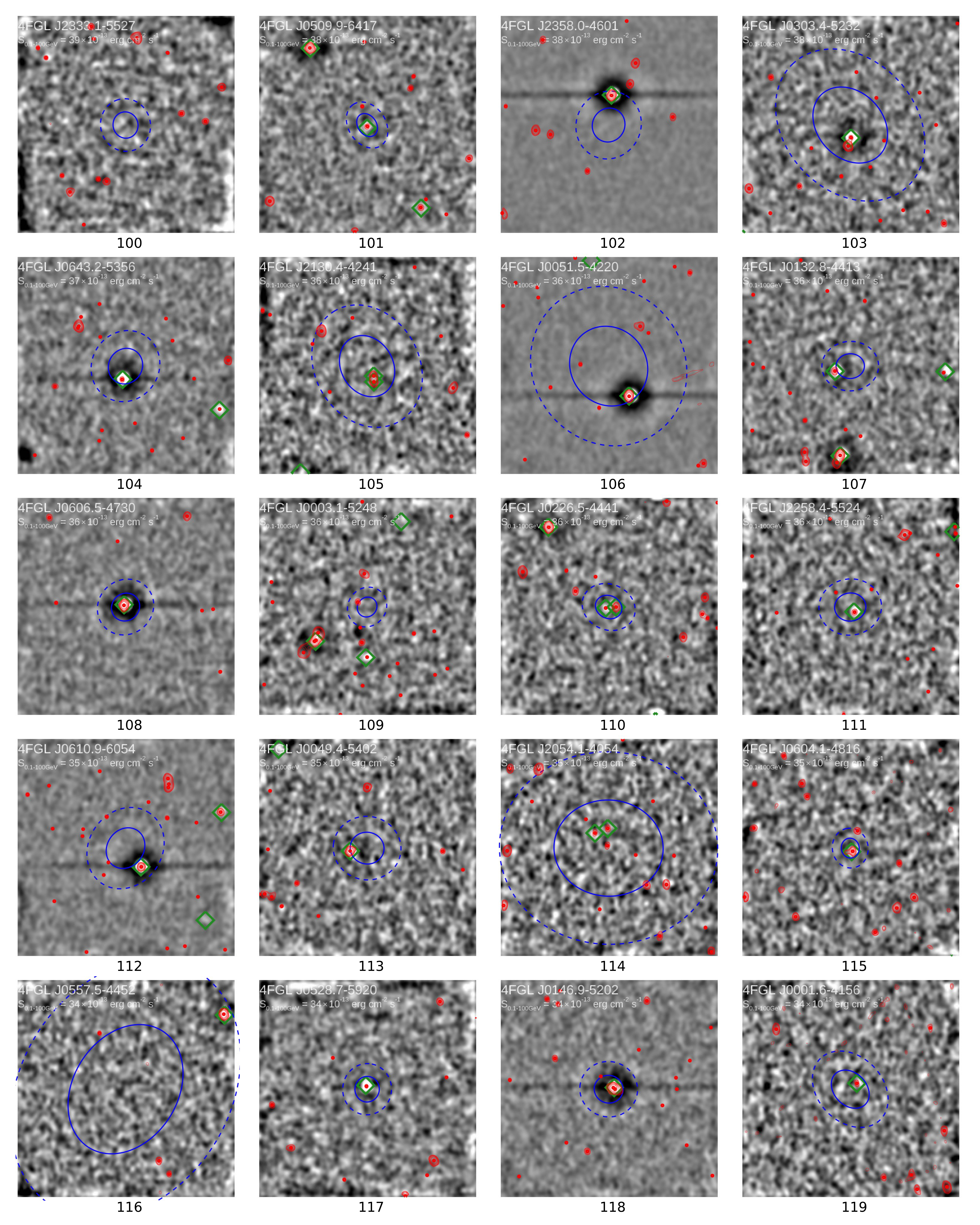}
\figsetgrpnote{SPT thumbnails of 282 4FGL sources. The full version can also be downloaded in \url{https://github.com/lizhong-phys/4FGL-SPT.git}.
}
\figsetgrpend

\figsetgrpstart
\figsetgrpnum{12.7}
\figsetgrptitle{7}
\figsetplot{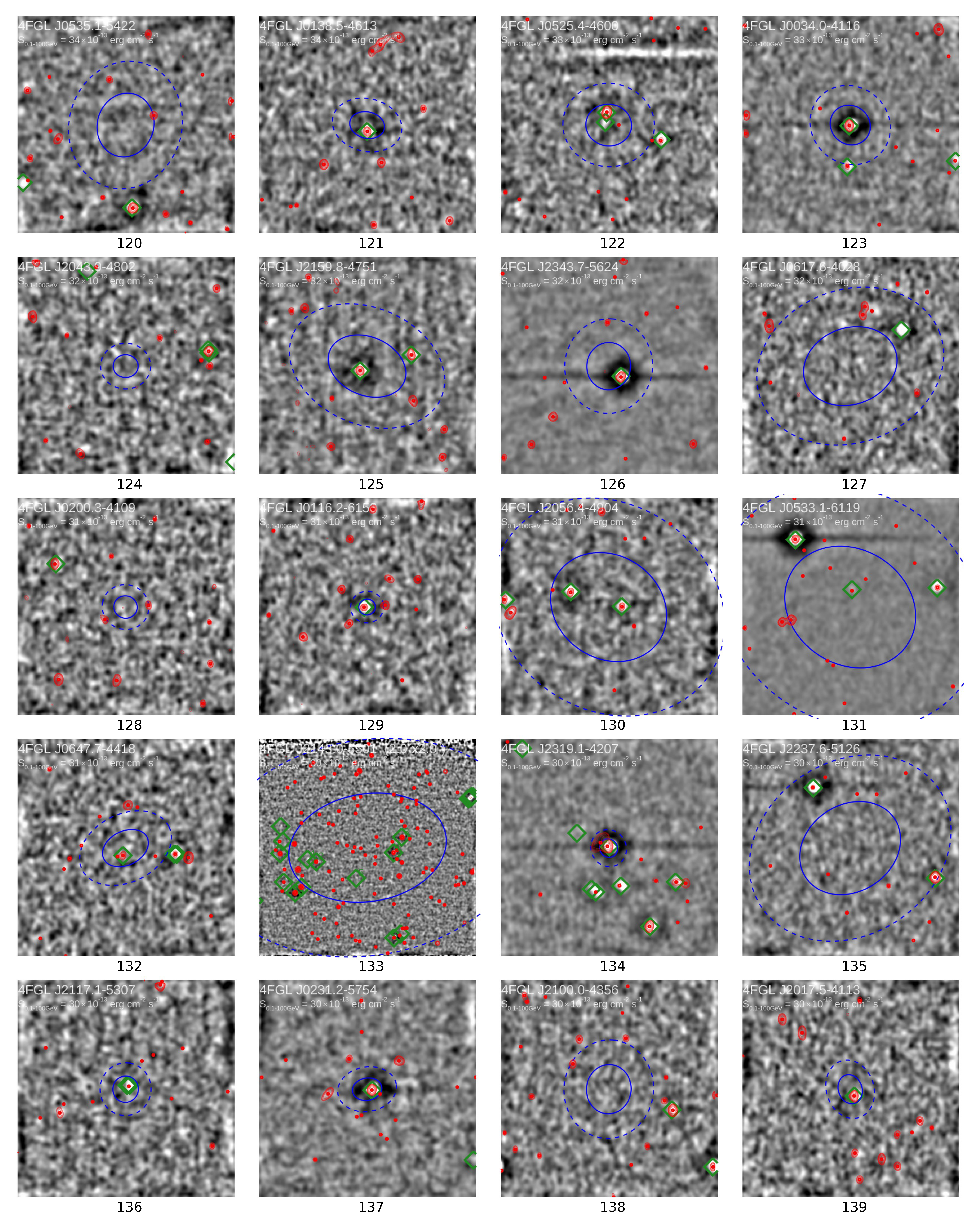}
\figsetgrpnote{SPT thumbnails of 282 4FGL sources. The full version can also be downloaded in \url{https://github.com/lizhong-phys/4FGL-SPT.git}.
}
\figsetgrpend

%
%
%
%
%
%
%

\figsetend

\begin{figure}
\figurenum{12}
\plotone{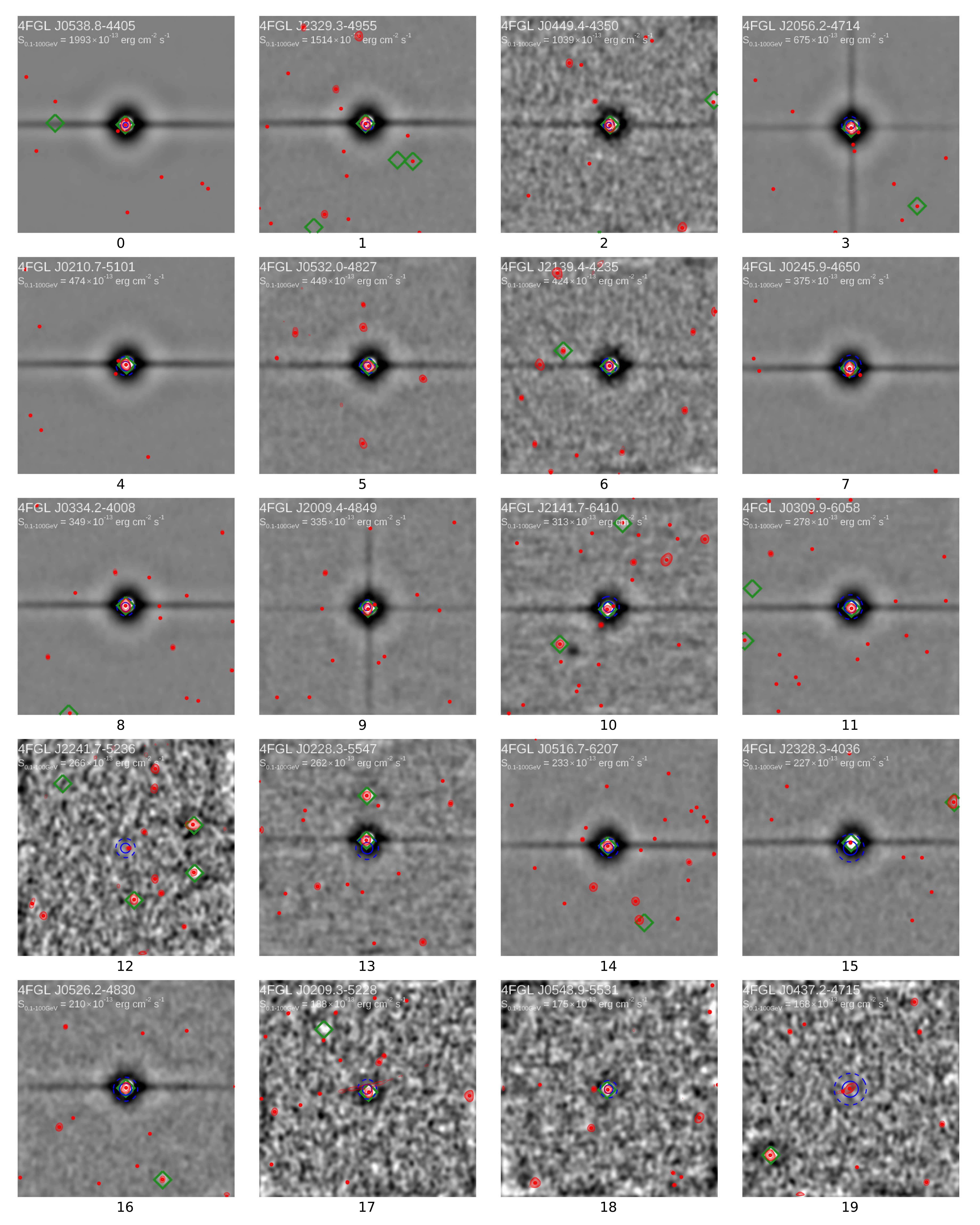}
\caption{SPT thumbnails of 282 4FGL sources. The full version can also be downloaded in \url{https://github.com/lizhong-phys/4FGL-SPT.git}.
}
\end{figure}

\end{appendices}

\end{document}